\documentclass{nature}
\usepackage{amsmath,amsthm,amssymb}
\usepackage{graphicx}
\usepackage[dvipsnames]{xcolor}

\usepackage[normalem]{ulem}

\usepackage{physics}					% various macros (derivatives, etc.)
\usepackage{bm}						% bold math
\usepackage{xfrac}

\usepackage{caption}
\DeclareCaptionLabelSeparator{bar}{ $|$ }% or $\vert$
\makeatletter
\let\saved@includegraphics\includegraphics
\AtBeginDocument{\let\includegraphics\saved@includegraphics}
\renewenvironment*{figure}{\@float{figure}}{\end@float}

\makeatother

\DeclareMathOperator {\e}{e}				% e for exponential should be roman

\title{Exponentially faster cooling in a colloidal system}
 
\author{Avinash Kumar$^{1}$ \& John Bechhoefer$^{1}$}

\begin{document}
\maketitle

\begin{affiliations}
 \item Dept. of Physics, Simon Fraser University, 8888 University Dr., Burnaby, BC, V5A 1S6, Canada.
\end{affiliations}

\begin{abstract}
Since the temperature of an object that cools decreases as it relaxes to thermal equilibrium, naively a hot object should take longer to cool than a warm one.  Yet, some 2300 years ago, Aristotle observed that ``to cool hot water quickly, begin by putting it in the sun"\cite{Webster1923aristotle, jeng2006mpemba}.  In the 1960s, this counterintuitive phenomenon was rediscovered as the statement that ``hot water can freeze faster than cold water'' and has become known as the ``Mpemba effect"\cite{mpemba1969cool}; it has since been the subject of much experimental investigation\cite{woj1988, Auerbach1995supercooling, vynnycky2012axisymmetric, vynnycky2015convection, burridge2016questioning} and some controversy\cite{burridge2016questioning,katz2017reply}.  While many  specific mechanisms have been proposed\cite{mirabedin2017, vynnycky2010evaporative, vynnycky2012axisymmetric, vynnycky2015convection, katz2009hot, esposito2008mpemba, zhang2014hydrogen, jin2015mechanisms, Tao2016Hbonding}, no general consensus exists as to the underlying cause.   Here we demonstrate the Mpemba effect in a controlled setting, the thermal quench of a colloidal system immersed in water, which serves as a heat bath.  Our results are reproducible and agree quantitatively with calculations based on a recently proposed theoretical framework\cite{lu2017nonequilibrium}.  By carefully choosing parameters, we observe cooling that is exponentially faster than that observed using typical parameters, in accord with the recently predicted \textit{strong Mpemba} effect\cite{Marija2019robust}.  Our experiments give a physical picture of the generic conditions needed to accelerate heat removal and relaxation to thermal equilibrium and support the idea that the Mpemba effect is not simply a scientific curiosity concerning how water freezes into ice---one of the many anomalous features of water\cite{sun16}---but rather the prototype for a wide range of \textit{anomalous relaxation} phenomena of broad technological significance.
\end{abstract}

That an initially hot object might cool more quickly than an initially warm object seems impossible, because our intuitions tend to be shaped by systems that remain at or near thermal equilibrium.  If an object is cooled \textit{slowly}, its time-dependent state is well characterized by a temperature, and a hot object cannot cool without passing through all intermediate temperatures.  Nonetheless, when rapidly quenched by placing a system in contact with a cold bath, the Mpemba effect is often observed in settings where a phase transition occurs\cite{burridge2016questioning, katz2017reply}.  There is still no widely accepted specific mechanism to explain the observations in water, where factors ranging from evaporation\cite{mirabedin2017, vynnycky2010evaporative}, convection currents\cite{vynnycky2012axisymmetric, vynnycky2015convection}, dissolved gases\cite{katz2009hot, woj1988}, supercooling\cite{esposito2008mpemba, Auerbach1995supercooling}, and hydrogen bonding\cite{zhang2014hydrogen, jin2015mechanisms, Tao2016Hbonding} have all been suggested.  Analogues of anomalous cooling behaviour reported in water have been reported in other systems with phase transitions, including magnetic systems\cite{chaddah2010overtaking}, clathrate hydrates\cite{ahn2016experimental}, and polymers\cite{hu18}.  In addition, numerical simulations predict Mpemba-like behaviour in granular fluids\cite{lasanta2017hotter,torrente2019large}, spin glasses\cite{baity2019mpemba}, nanotube resonators\cite{greaney2011mpemba}, quantum systems\cite{Fabrizio2019phase}, and cold gases\cite{keller2018quenches}. Here we provide clear experimental evidence for the  Mpemba effect in a colloidal system that lacks a phase transition.  Our results are the first to agree quantitatively with theoretical predictions giving a  \textit{general} explanation for the Mpemba effect\cite{lu2017nonequilibrium}, and we take advantage of our understanding of the underlying physics to achieve cooling times that are exponentially faster than the time to cool under typical initial conditions.

\section*{Definition of the Mpemba effect}

Past investigations of the Mpemba effect have suffered from vague, mutually inconsistent definitions.  Here we define the Mpemba effect in terms of three temperatures $T_\text{h}~>~T_\text{w}~>~T_\text{c}$, for which the time $t_\text{h}$ to cool a system from a hot to a cold state is shorter than the time $t_\text{w}$ to cool it from an intermediate warm state to the same cold state. In the above definition, ``hot" describes an initial state that is at thermal equilibrium at temperature $T_\text{h}$, while ``warm" describes an initial state at thermal equilibrium at temperature $T_\text{w}$.  The cold temperature  $T_\text{c}=T_\text{b}$ is that of the thermal bath of water and is the identical final state for all initial conditions studied.  All terms in our definition are unambiguous:  the only ingredients are the equilibrium start and end states, characterized by the usual notion of temperature, and the time it takes to go from one state to another.  By contrast, previous definitions of the Mpemba effect have been based on criteria such as the ``time to start freezing"\cite{mpemba1969cool,katz2017reply}, which is hard to reproduce because of sensitivity to details of sample preparation\cite{jeng2006mpemba}, including impurities in the water, cleanliness of the sample container, and number of times heated.

\section*{Experimental approach}

\begin{figure}[ht!]
\centering
\includegraphics[width=1\linewidth]{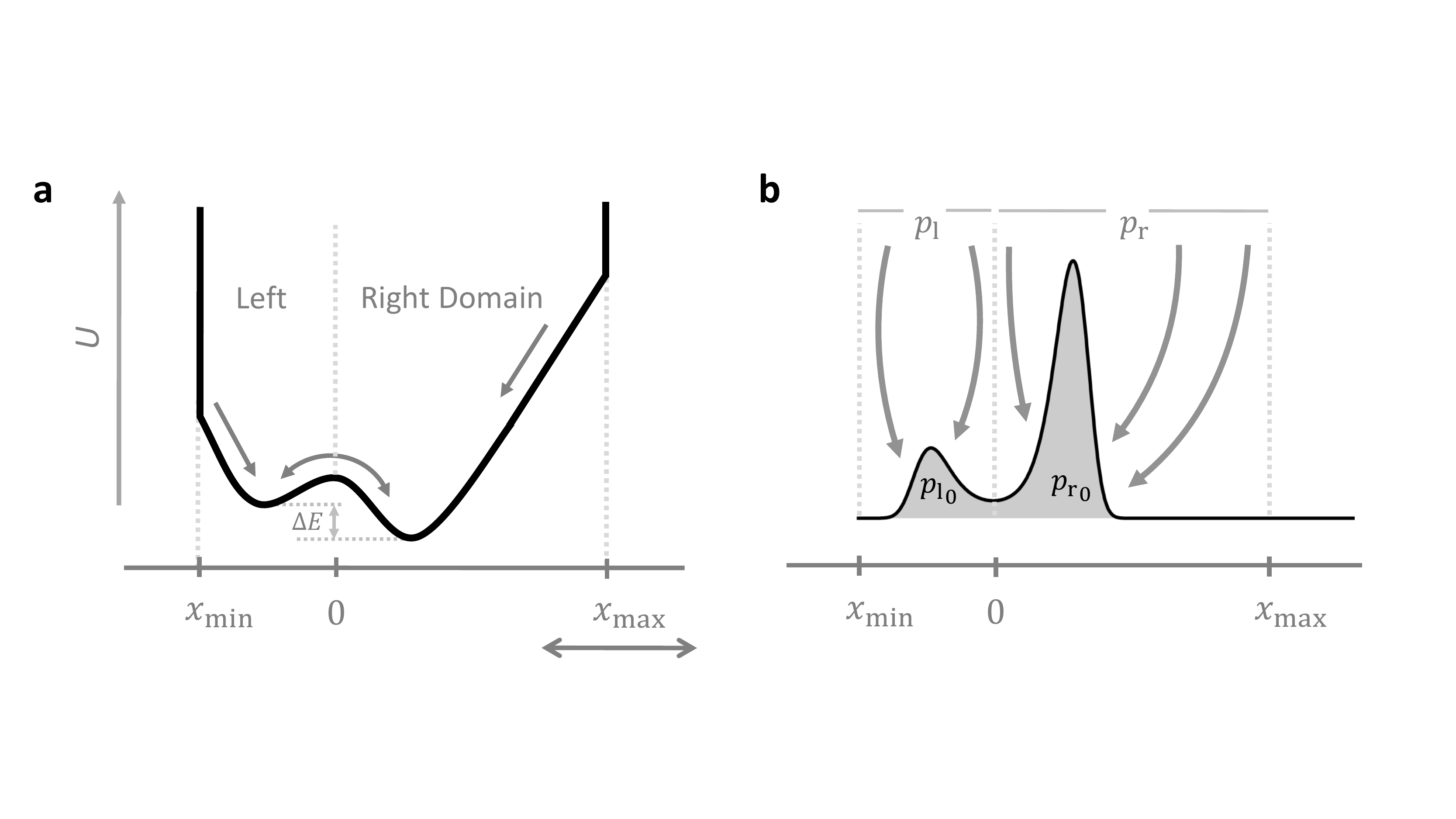}
\caption{\textbf{Schematic diagram of the energy landscape and Boltzmann distribution for the Mpemba effect.}  \textbf{a}, The solid black line represents the energy landscape $U(x)$, set asymmetrically within the box $\left[x_\text{min}, x_\text{max}\right]$ with infinite potential walls at the domain boundaries.  The asymmetry coefficient $\alpha \equiv |x_\text{max}/x_\text{min}|$, where $x_\text{max}$ is changed to vary $\alpha$.  The outer slopes of the potential correspond to the maximum force that can be exerted by the optical tweezers. The difference in energy between metastable and stable states is $\Delta E$. \textbf{b}, The Boltzmann distribution at the bath temperature, $\pi(x;T_\text{b})$. The interval $p_\text{r}$ is the probability for a particle to initially be in the right domain for $T_\text{h}\to \infty$.  At the bath temperature, the probability to be in the ground state (right well) is $p_{\text{r}_{0}}$.  We find that the Mpemba effect is strongest for $p_r \approx p_{\text{r}_{0}}$, a condition that allows the probability contained in the initial ``basin of attraction" to drain directly to the ground state.  Similarly, for the left well, $p_\text{l}  \approx p_{\text{l}_{0}}$, with  $p_\text{l} = 1- p_\text{r}$ and $p_{\text{l}_{0}} = 1- p_{\text{r}_{0}}$.}
\label{fig:potential_cartoon}
\end{figure}

In each experimental trial, a single Brownian particle diffuses in water, subject to forces from a carefully shaped potential (Fig.~\ref{fig:potential_cartoon}).  The potential is created using the force supplied by optical tweezers, as directed by a feedback loop; it is thus a \textit{virtual potential} (Methods).  Its form $U(x)$ consists of a tilted double well whose outer edges have a slope that saturates at a magnitude corresponding to the maximum force $F_\text{max}$ that the optical tweezers can exert.  The tilted double well creates a bistable potential with two macrostates:  the shallow left well corresponds to a metastable macrostate and the deep right well to a stable macrostate.  The linear parts of the potential provide direct kinetic paths towards the minima, and the barrier allows spontaneous hopping between the wells.  Because spatial dimensions are small and the energy barrier low, the bead can rapidly equilibrate with the bath ($\lesssim 0.1$ s).  We can then easily carry out several thousand trials, forming a statistical ensemble from which accurate measurements of both equilibrium and nonequilibrium states are possible.  As we will see below, to observe the Mpemba effect, we should place the potential asymmetrically between the potential boundaries, $x_\text{min}$ and $x_\text{max}$, which determine the region in space explored by the particle at high temperatures.  The parameter $\alpha \equiv | x_\text{max} / x_\text{min}|$ defines the degree of asymmetry within the domain ``box." In our experiments, we change $\alpha$ by varying $x_\text{max}$ while fixing $x_\text{min}$.

In our experiments, the particle is always in contact with water at temperature $T_\text{b}$; however, the initial state of the system is drawn from a Boltzmann distribution at a \textit{higher} initial temperature.  All temperatures are measured relative to the bath temperature $T_\text{b}$, and all energies are scaled by $k_\text{B}T_\text{b}$. After an effectively instantaneous quench at $t=0$, the particle position evolves according to the imposed virtual potential $U(x)$ under thermal environment fluctuations for 60 ms.  This protocol is repeated $N = 1000$ times, with the resulting data used to create a statistical ensemble from which we estimate the state of the system every 10 \textmu s.  

\begin{figure}[ht!]
\centering
\includegraphics{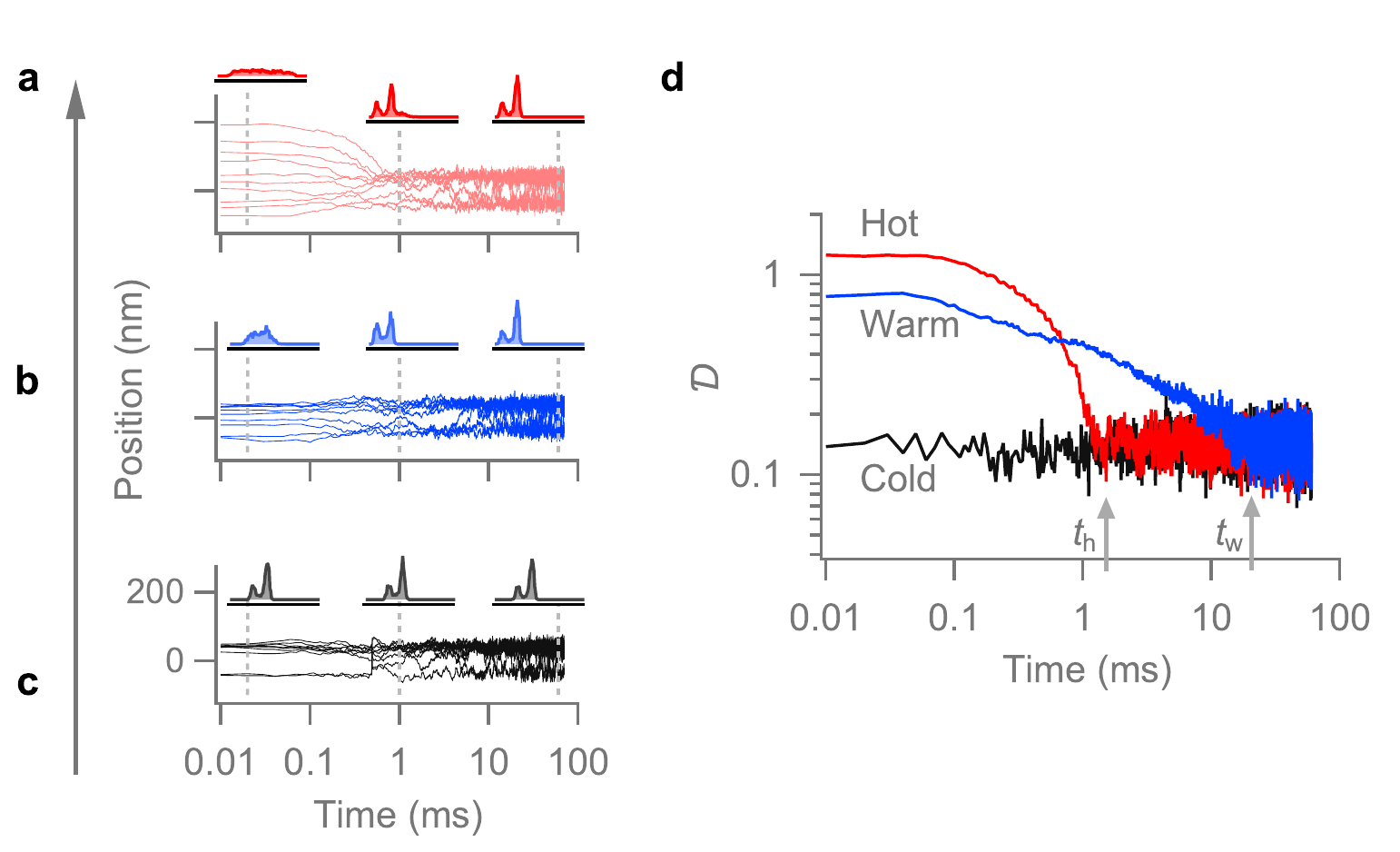}
\caption{\textbf{Dynamics of system relaxation to equilibrium.} \textbf{a}--\textbf{c}, Ten trajectories of a particle released from the equilibrium distributions at hot (red), warm (blue), and cold (black) temperatures into the cold bath, with the evolving probability density $p(x,t)$ shown for three times (estimates based on 1000 trajectories).  \textbf{d}, The $L_1$ distances calculated for systems at three different temperatures ($T_\text{h} = 1000$, $T_\text{w} = 12$, and $T_\text{c} = 1$) from their respective time traces. The initially hot system starts a greater distance from equilibrium than the initially warm system but equilibrates first ($t_\text{h} < t_\text{w}$), illustrating the Mpemba effect.  The cold distance plot is a control experiment where the particle is released from the equilibrium distribution at the bath's temperature.  At each time, the distribution fluctuates from the average by an always-positive distance. There are $N = 1000$ runs per initial temperature; asymmetry coefficient $\alpha =3$.}
\label{fig:Mp_Traj}
\end{figure}

Figure~\ref{fig:Mp_Traj}a--c shows example time traces of evolution in the potential $U(x)$.  From the time traces, we form frequency estimates of the probability density function $p(x,t)$ that records the system state as it evolves between the initial state $p(x,0) = \pi(x;T_\text{initial}) \propto \exp [-U(x)/k_\text{B}T_\text{initial}]$ and the final state at equilibrium with the bath, characterized by $\pi(x;T_\text{b})$.

At intermediate times while the system is relaxing, the dynamical state $p(x,t)$ does not in general have the form of a Boltzmann distribution for the potential $U(x)$ at any temperature; nevertheless, we can define a scalar quantity\cite{lebowitz1957irreversible,risken89}  that measures the ``distance" $\mathcal{D}$ between $p(x,t)$ and the Boltzmann distribution in equilibrium with the bath, $\pi(x;T_\text{b})$.  For simplicity, we choose an $L_1$ measure of distance, but any measure that is monotonic with $T_\text{initial}$ will also work (Methods).  As shown in Fig.~\ref{fig:Mp_Traj}d, we can use the $L_1$ distance curve to determine the time at which the system reaches equilibrium ($\mathcal{D} \approx 0$, within noise levels---black curve; see Methods).

\section*{Observation of the Mpemba effect in asymmetric domains}

\begin{figure}[ht!]
\centering
\includegraphics{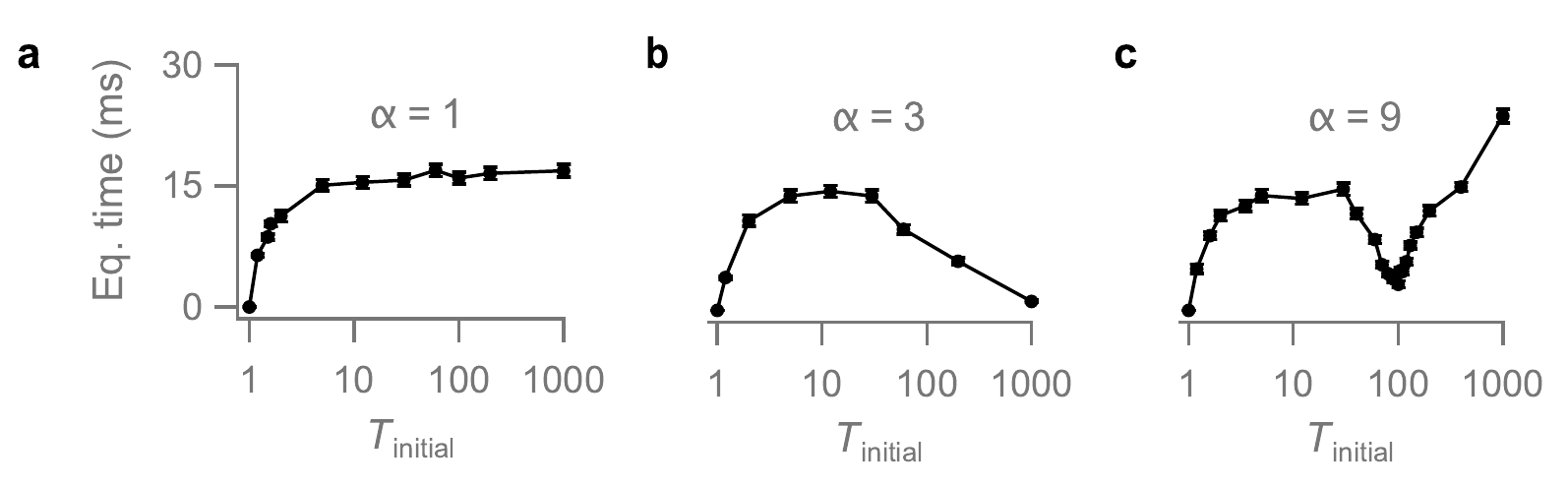}
\caption{\textbf{Equilibration time as a function of initial system temperature.} \textbf{a}--\textbf{c}, Solid markers  are the equilibration times for asymmetry coefficients $\alpha =1, 3, 9$.  Regions of each plot with negative slope indicate the Mpemba effect. The error bars represent standard deviations calculated using Eq.~\ref{Eq:SD_t_eq} (Methods).}
\label{eq_time}
\end{figure}

To determine how the Mpemba effect depends on the shape of the potential, we first place the double-well potential in a symmetric box ($\alpha =1$).  Figure \ref{eq_time}a shows the measured times to reach equilibrium for systems that start at different initial temperatures.  The equilibration time increases sharply and saturates at high temperatures, where the initial probability distribution is nearly uniform.  Since the equilibration time monotonically increases with initial temperature, there is no Mpemba effect.  

The situation changes qualitatively when $x_\text{max}$ is increased and the box becomes asymmetric. For $\alpha =3$ (Fig.~\ref{eq_time}b), the equilibration time increases initially but then decreases rapidly for higher temperatures ($T_\text{initial} > 10$), indicating the Mpemba effect.  For $\alpha =9$ (Fig.~\ref{eq_time}c), the equilibration time decreases at intermediate temperatures, where the Mpemba effect is observed, but increases again at very high temperatures.

\begin{figure}[ht!]
\centering
\includegraphics[width=1\linewidth]{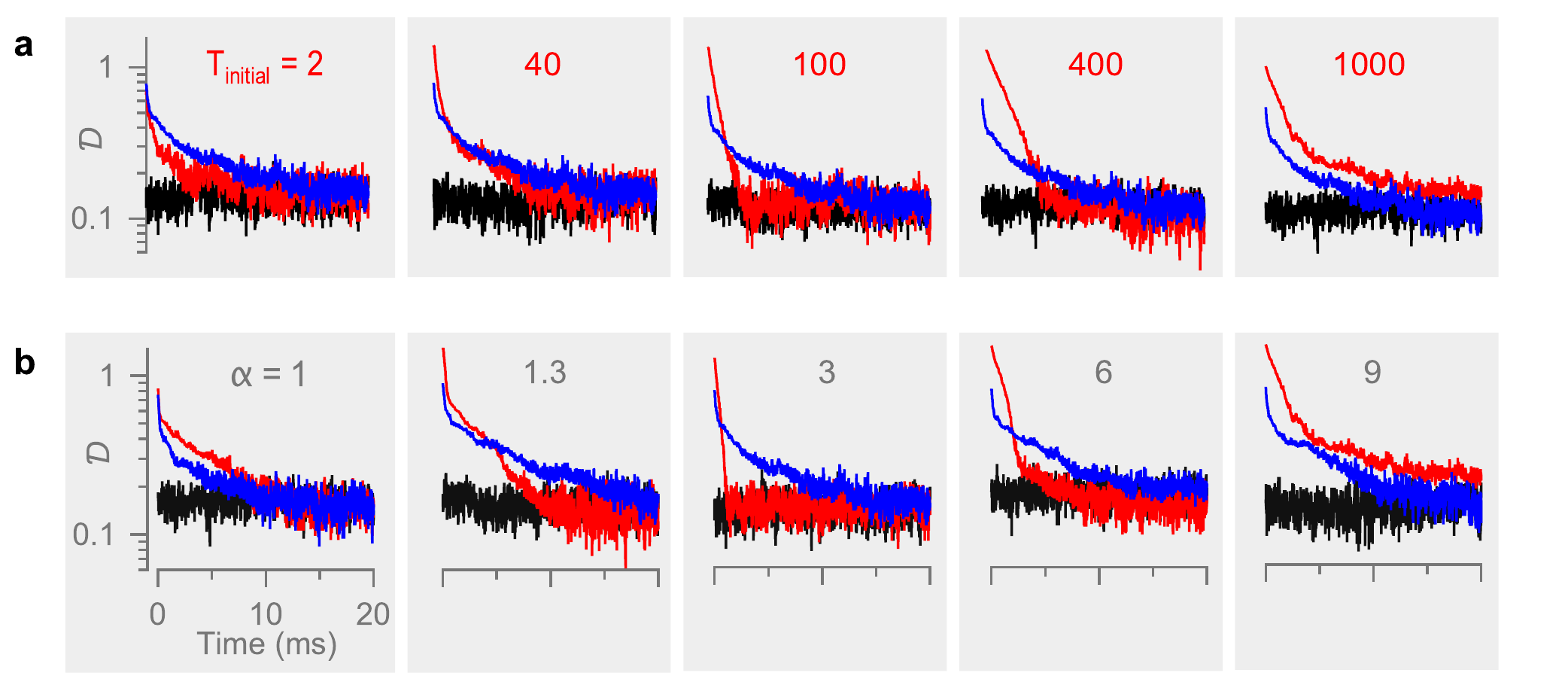}
\caption{\textbf{Controlling relaxation times.}   \textbf{a}, $L_1$ distance for systems with initial temperatures $T_\text{c}=1$ (black), $T_\text{w}=12$ (blue), and $T_\text{h} = \{2, 40, 100, 400, 1000 \}$ (red), with $\alpha = 9$.  At low and high $T_\text{h}$, no Mpemba effect is observed.  At intermediate $T_\text{h}$, the distance curves cross, indicating a more rapid cooling of the hot system.  \textbf{b}, $L_1$ distance for domain asymmetries $\alpha = \{1, 1.3, 3, 6, 9 \}$, for $T_\text{c}=1$ (red), $T_\text{w}=12$ (blue), and $T_\text{h}=1000$ (red). The Mpemba effect is observed for intermediate asymmetry. The control experiment is repeated for each measurement.}
\label{fig:Mp_temp_alpha}
\end{figure}

To understand the different equilibration-time curves in Fig.~\ref{eq_time}, we should examine more closely the distance curves $\mathcal{D}(t)$, which summarize the relaxation of the system to thermal equilibrium.  Figure~\ref{fig:Mp_temp_alpha}a shows data for $\alpha = 9$, corresponding to the curve in Fig.~\ref{eq_time}c.  On the semilog plots, straight lines represent exponential decay. For $T_\text{initial} = 100$, the Mpemba effect is particularly clear, and the system appears to relax to equilibrium as a single exponential. For other initial temperatures, the relaxation seems to involve multiple exponential relaxation processes.

\section*{Analysis based on eigenfunction expansion}

To interpret the dynamical behaviour of $\mathcal{D}(t)$, we apply a recent approach\cite{lu2017nonequilibrium} that connects the Mpemba effect to an eigenvalue expansion.  The underlying probability density $p(x,t)$ can be expressed as an infinite sum of eigenfunctions of the Fokker-Planck equation (FPE), which governs the evolution of $p(x,t)$.  The $k^{\text{th}}$ eigenfunction $v_k(x; \alpha, T_\text{b})$ is a spatial function that depends on the form of the potential $U(x)$, including the asymmetry coefficient $\alpha$, and the bath temperature $T_\text{b}$.  The contribution of eigenfunction $v_k$ decays exponentially, at a rate $\exp(-\lambda_k t)$, where the eigenvalues $\lambda_k \geq 0$ are ordered so that $0=\lambda_1 < \lambda_2 < \cdots$.  At long times, the theory then predicts (Methods, Eq.~\ref{Eq:FPsolution}) that the density function is dominated by the first two terms of the infinite series:

\begin{align}
	p(x,t) ~\approx~ \pi(x;T_\text{b}) + {a_2(\alpha,\,T_\text{initial}) 
	\e^{-\lambda_2 t}v_2(x; \alpha,T_\text{b})} \,,
\label{p2}
\end{align}

where the coefficient $a_2(\alpha,\,T_\text{initial})$ is a real number that depends on the initial temperature and the potential energy. It captures the ``overlap" between the initial system state and the second left eigenfunction (Methods, Eq.~\ref{eq:a2}).

Equation~\eqref{p2} has several consequences (Methods):
\begin{itemize}
\item The equilibration time of an initial state depends on its $a_2$ coefficient.  
\item The difference in equilibration times $t_\text{w}-t_\text{h}$ is independent of the noise level of $\mathcal{D}(t)$.  
\item The magnitude of the $a_2$ coefficient may be extracted from $\mathcal{D}(t)$. 
\item For $a_2 = 0$, the system reaches equilibrium at an exponentially faster rate (decay dominated by $\lambda_3 > \lambda_2$). 
\item The Mpemba effect correlates with the condition\cite{lu2017nonequilibrium} that $|a_2(\alpha, T_\text{h})| < |a_2(\alpha, T_\text{w})|$.  
\end{itemize}
The last point implies that the Mpemba effect occurs over a range of initial temperatures for which $|a_2(\alpha,T_\text{initial})|$ decreases as $T_\text{initial}$ increases.  

\begin{figure}[ht!]
\centering
\includegraphics[width = 1\linewidth]{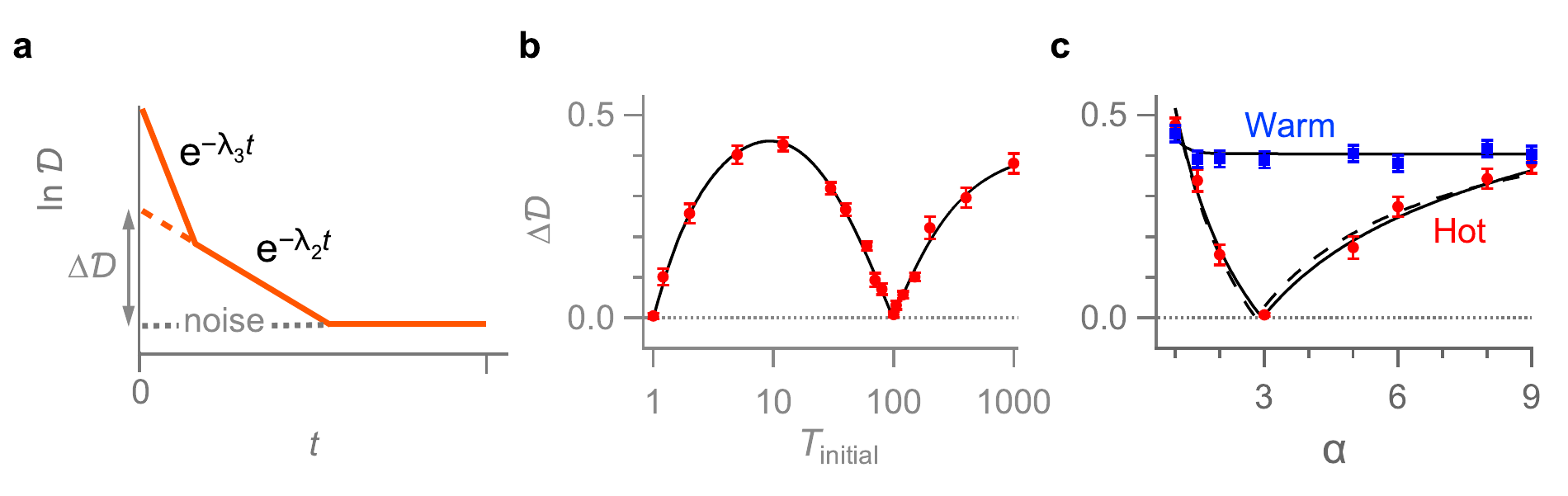}
\caption{\textbf{Measurements of \bm{$\Delta \mathcal{D}$}.}  \textbf{a}, $\Delta \mathcal{D}$ is measured by extrapolating the long-time limit of the logarithm of the $\mathcal{D}$ curve back to the quench time ($t=0$).  \textbf{b}, Markers are $\Delta \mathcal{D}$ measurements based on distance plots for different initial temperatures for $\alpha = 9$.  \textbf{c}, Red and blue markers denote $\Delta \mathcal{D}$ measurements based on distance plots of the hot ($T_\text{h}=1000$) and warm ($T_\text{w}=12$) systems for different asymmetry coefficients. Solid lines in \textbf{b}, \textbf{c} are based on the calculated $|a_2(\alpha,\, T_\text{initial})|$ coefficient multiplied by an experimentally determined scaling factor.  Dashed line in \textbf{c} shows fit based on Eqs.~\eqref{eq:pdef}, \eqref{linear approx}.  The error bars represent one standard deviation and are calculated from the fits.}
\label{fig:deltaD}
\end{figure}

Following these points, we analyze the $\mathcal{D}(t)$ curves by extracting from them a quantity $\Delta \mathcal{D}$ that is sketched in Fig.~\ref{fig:deltaD}a. For $a_2 \neq 0$, it may be estimated by globally fitting a single exponential to the long-time asymptotic regimes of all the $T_\text{h}$ decays in Fig.~\ref{fig:Mp_temp_alpha} (Methods, Eq.~\ref{Eq:a2vsdelD}), extrapolating back in time to find the intercept at $t = 0$, and subtracting the noise level $\sigma_\mathcal{D}$ resulting from finite sampling (Methods, Eq.~\ref{Eq: L1error}).  We can then show that $\Delta \mathcal{D}~\propto~|a_2(\alpha,\,T_\text{initial})|$.  The proportionality constant depends on the $v_2$ eigenfunction and may be calculated given the potential $U(x)$ and the bath temperature $T_\text{b}$ (Methods, Eq.~\ref{Eq:delD}).

Carrying out the analysis sketched above, we extract  $a_2(\alpha, T_\text{initial})$ and confirm that the Mp-emba effect is indeed associated with an $a_2$ that decreases with the initial temperature (Fig.~\ref{fig:deltaD}b).  We fit the measured values of $\Delta \mathcal{D}(\alpha,T_\text{initial})$ with the numerical result from the FPE to estimate the scaling factor multiplying the $|a_2(\alpha,T_\text{initial})|$ coefficients. The scaling factor from the fit, $0.96 \pm 0.03$, agrees with the numerical value $\approx 0.967$ calculated using the FPE and its numerically determined eigenfunctions. The variation of $\Delta \mathcal{D}$ (and thus, $a_2$) with temperature for this fixed $\alpha$ shows non-monotonic behaviour that also reflects the presence of the Mpemba effect. However, the eigenfunction analysis does not itself explain \textit{why} the $a_2$ coefficient has a non-monotonic dependence on $T_\text{initial}$.  

\section*{Strong Mpemba effect}

To gain more physical insight, we conducted further experiments probing the Mpemba effect at fixed temperatures but variable asymmetry.  In particular, we fixed the hot initial temperature $T_\text{h}=1000$, which is so high that the initial probability density $p(x,0) = \pi(x; T_\text{h})$ is approximately a uniform distribution over the domain $(x_\text{min},x_\text{max})$.  Figure~\ref{fig:Mp_temp_alpha}b shows distance plots for different $\alpha$, for hot and warm initial temperatures $T_\text{h} = 1000$, $T_\text{w} = 12$.  As the asymmetry varies from $\alpha =1$ to $\alpha=9$, we see the same sequence of normal, anomalous (Mpemba), and normal relaxations to thermal equilibrium that we saw in Fig.~\ref{fig:Mp_temp_alpha}a, where $\alpha$ was fixed and $T_\text{initial}$ was varied.

Figure ~\ref{fig:deltaD}c shows that the $\Delta \mathcal{D}$ values calculated from the experimental data presented in Fig.~\ref{fig:Mp_temp_alpha}b are linearly proportional to the $a_2$ coefficient. We first notice that the value of the $a_2$ coefficient for the warm system is roughly constant, as increasing the asymmetry does not change its initial state.  The behaviour of the $a_2$ coefficient for the hot system is more complicated.  For small asymmetry, $|a_2(\alpha,\,T_\text{w})| < |a_2(\alpha,\,T_\text{h})|$, and the warm system cools down faster; i.e., $t_\text{w} < t_\text{h}$ (Eq.~\ref{time_diff}).  This corresponds to normal cooling.  For larger asymmetries, the situation is reversed, and we observe the Mpemba effect.  For the special asymmetry value $\alpha$ = 3, the $|a_2(\alpha,\,T_\text{h})|$ coefficient vanishes.  Such a situation corresponds to the recently identified strong Mpemba effect\cite{Marija2019robust} and implies an exponential speed-up of the cooling process.    

In the limit of large $T_\text{h}$, it is easy to understand this normal-anomalous-normal sequence of relaxation behaviour.  Because the initial probability density at $T_\text{h}$ is approximately uniform, we can approximate the relative probability $p_\text{r}$ to be in the right-hand domain $(0,x_\text{max})$ as

\begin{align}
 	p_\text{r} = \frac{|x_\text{max}|}{|x_\text{min}|+|x_\text{max}|}  = \frac{\alpha}{1+\alpha} \,.
\label{eq:pdef}
\end{align}

We can refer to this subset of initial conditions as the \textit{ground-state basin} because it constitutes the states that, in the absence of thermal fluctuations, would flow into the well corresponding to the stable state.  Similarly, the metastable state attracts the initial conditions $(x_\text{min},0)$, which may be termed the \textit{metastable-state basin}.  On the one hand, when the particle is released from its initial position to evolve under the influence of the potential, it rapidly moves to one of the two wells.  Thus, after a fast transient, we expect the probability to be in the ground-state well to be $\approx p_\text{r}$.  On the other hand, the measured probability for a system in thermal equilibrium (Fig.~\ref{fig:potential_cartoon}b) to occupy the ground state is $p_{\text{r}_{0}} \approx 0.7$.  If the asymmetry $\alpha$ is chosen so that $p_\text{r} = p_{\text{r}_{0}}$, then the system will be in equilibrium after this initial transient.  But for any other $\alpha$, there will be a mismatch and $p_\text{r} \neq p_{\text{r}_{0}}$.  The system will then relax to equilibrium by thermal hops over the barrier.  This process is slowed by the Kramers factor $\exp (E_\text{barrier} / {k_\text{B}T_\text{b}}) \approx 7$ in our system, implying a longer time to reach equilibrium.

The above argument leads to a simple prediction for the asymmetry dependence of the $a_2$ coefficient for the hot system in Fig.~\ref{fig:deltaD}c and hence for $\Delta \mathcal{D}$.  If the dynamic and the reference probabilities are close, we can approximate their difference using a Taylor expansion,

\begin{align}
	\Delta \mathcal{D}(\alpha,T) \propto |p_\text{r}-p_0| \,.
\label{linear approx}
\end{align}

We then fit to the data shown in Fig.~\ref{fig:deltaD}c (dashed line).  The fit agrees well with the experimental observations and with a numerical calculation based on the FPE eigenfunctions (solid line).

\section*{Discussion}

We have experimentally demonstrated the Mpemba effect.  Our study gives insight into a long-standing problem and represents the first case where quantitative agreement between a predictive theory and experiment is observed.

The significance of observing the Mpemba effect in a colloidal system is twofold:  First, simplicity brings clarity.  The agreement shown with a simple theory\cite{lu2017nonequilibrium} based on eigenfunction expansions of the FPE contrasts with the more complicated, yet inconclusive analyses of the ice-water system\cite{katz2009hot, esposito2008mpemba, vynnycky2012axisymmetric, vynnycky2015convection, burridge2016questioning, katz2017reply, mirabedin2017, vynnycky2010evaporative, woj1988, Auerbach1995supercooling, zhang2014hydrogen, jin2015mechanisms, Tao2016Hbonding}.  More constructively, the physical insights gained from the study of a simple system may guide future investigations of more complicated systems.  For example, while many authors have asserted that freezing plays an essential role\cite{katz2017reply} in the water experiments, there is no phase transition in the experiments reported here;  however, the potential does have a metastable well, suggesting the need for a region in state space that can act as a temporary trap for dynamical trajectories en route to equilibrium.  Note that while our state space is a one-dimensional space of positions, our arguments apply to higher-dimensional spaces with multiple macrostates.  Some of the dimensions may correspond to internal degrees of freedom.  What is important is that the volumes in state space of the  basins of attraction of the initial system state nearly match the probabilities of the macrostates for the bath distribution (Fig.~\ref{fig:potential_cartoon}b).

The second significance of the colloidal experiments is to show that the ice-water system is not unique.  The analysis used here\cite{lu2017nonequilibrium} constitutes a \textit{general} mechanism for anomalous relaxation phenomena.  The situation is analogous to that of phase transitions, where general physical theories (mean-field and Landau theories, renormalization group)\cite{goldenfeld92} contrast with theories for specific cases such as the ice-water transition.  Detailed theories for specific systems can account for important phenomena in a given system, for example how additives increase the attainable supercooling in water and help insects survive sub-freezing temperatures\cite{debenedetti97}, while general theories such as we have applied suggest how similar behavior can arise in a wide variety of settings and materials.

Here we have used our understanding of the phenomenology of the Mpemba effect to identify special combinations of experimental parameters where the $a_2$ coefficient vanishes (strong Mpemba effect), which correspond to exponentially faster cooling.   More sophisticated time-dependent protocols can also be envisioned that steer dynamical trajectories to desired outcome states.  A recent theory along these lines shows that an initial cooling can actually speed up \textit{heating times} exponentially\cite{gal20}.  Indeed, searching for such an inverse Mpemba effect\cite{lu2017nonequilibrium} remains a tantalising experimental goal.  More broadly, thermal relaxation and heat removal remain important technological challenges.  For example, they limit the performance of microprocessors and other integrated circuits\cite{moore14}.  Engineering Mpemba-like effects into technologically relevant materials might offer new and important strategies to rapidly remove heat from localized sources.

\begin{addendum}
 \item We thank Oren Raz, Zhiyue Lu, Karel Proesmans, Rapha{\"e}l Ch{\'e}trite, Nancy Forde,  Steve Dodge, and Tushar Kanti Saha for helpful suggestions. We also thank Xiaoyi Su and especially Lisa Zhang\cite{zhang19}, who contributed to preliminary versions of the experiment.  This research work has been supported by Discovery and RTI Grants from the National Sciences and Engineering Research Council of Canada (NSERC).
 \item[Competing Interests] The authors declare that they have no competing financial interests.
 \item[Correspondence] Correspondence and requests for materials
should be addressed to J.B.~(email: johnb@\\sfu.ca).
 \item [Data and code availability] The data and the code that support the findings of this study are available from the corresponding author upon reasonable request.
\end{addendum}

\bibliographystyle{plain}

\bibliography{References}

%%%%%%%%%%%%%%%%Supplementary%%%%%%%%%%%%%%%%%

\newcommand{\beginsupplement}{%
        \setcounter{table}{0}
        \renewcommand{\thetable}{\arabic{table}}%
        \setcounter{equation}{0}
        \renewcommand{\theequation}{\arabic{equation}}%
        \setcounter{figure}{0}
       \renewcommand{\thefigure}{Extended Data Fig.~\arabic{figure}}
        \captionsetup[figure]{labelfont={bf},name={},labelsep=bar}
        }
        
\beginsupplement
\newpage

\section*{Methods}

\subsection{Setup \\ } 
\label{sec:setup}

The experimental setup needs to impose a carefully chosen energy landscape (potential) to the motion of a particle diffusing in water.  We use a recently designed feedback trap\cite{kumar2018} to impose a \textit{virtual} potential whose form we are free to choose\cite{cohen05,jun12}.  We note that the technology of feedback traps was crucial to our ability to carry out the experiments in this paper.  Two features are key:  First, we can freely and accurately choose the shape $U(x)$ of the imposed potential.  Second, the potential can vary on length scales well below the diffraction limit.  Here, the separation between wells corresponding to ground and metastable states was 80 nm.  Alternative techniques  have also been used to create custom potential shapes (e.g., holographic optical tweezers\cite{chupeau20} or time-shared optical tweezers\cite{berut12}), but these potentials have micron-scaled features, limited by diffraction (wavelength of light used for the tweezer). The $\approx 10$-fold decrease in length scales of feedback tweezers implies a 100-fold decrease in time scales.  Not only are measurements at comparable statistics 100 times faster, but the effects of temperature drifts on the equilibrium position of the trap become insignificant.

\begin{figure}[ht!]
\centering
\includegraphics{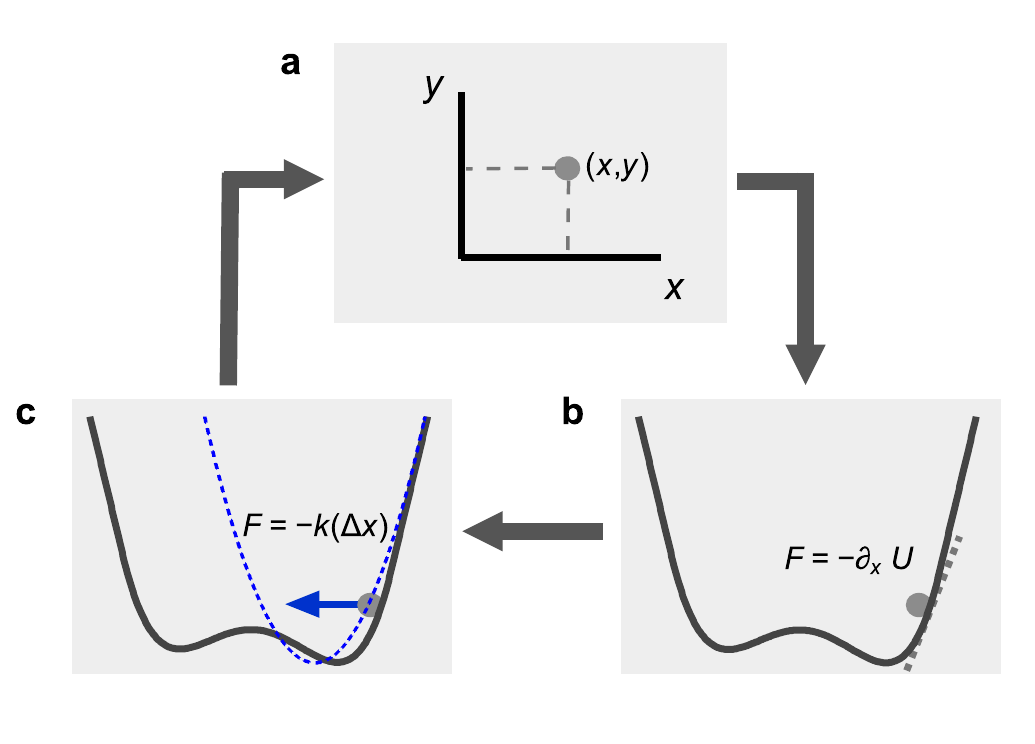}
\caption{\textbf{One cycle of a feedback trap.} \textbf{a}, Measure the particle position; \textbf{b}, Calculate force from the gradient of the imposed potential (black) based on the position; and \textbf{c}, Apply the force by shifting the harmonic trap centre (blue). The force applied is a linear restoring force with $k$ the stiffness of the harmonic trap and $\Delta x$ the imposed trap displacement.}
\label{fig:FB}
\end{figure}

Feedback traps operate by repeated cycles of a feedback loop based on (1) observation of the position of the particle, (2) calculation of the force based on its position in the user-defined potential, and (3) application of that force.  See ~\ref{fig:FB}.  In our design, the physical force is achieved by the application of an optical tweezer (OT).

\begin{figure}[t!]
\centering
\includegraphics[width=0.8\linewidth]{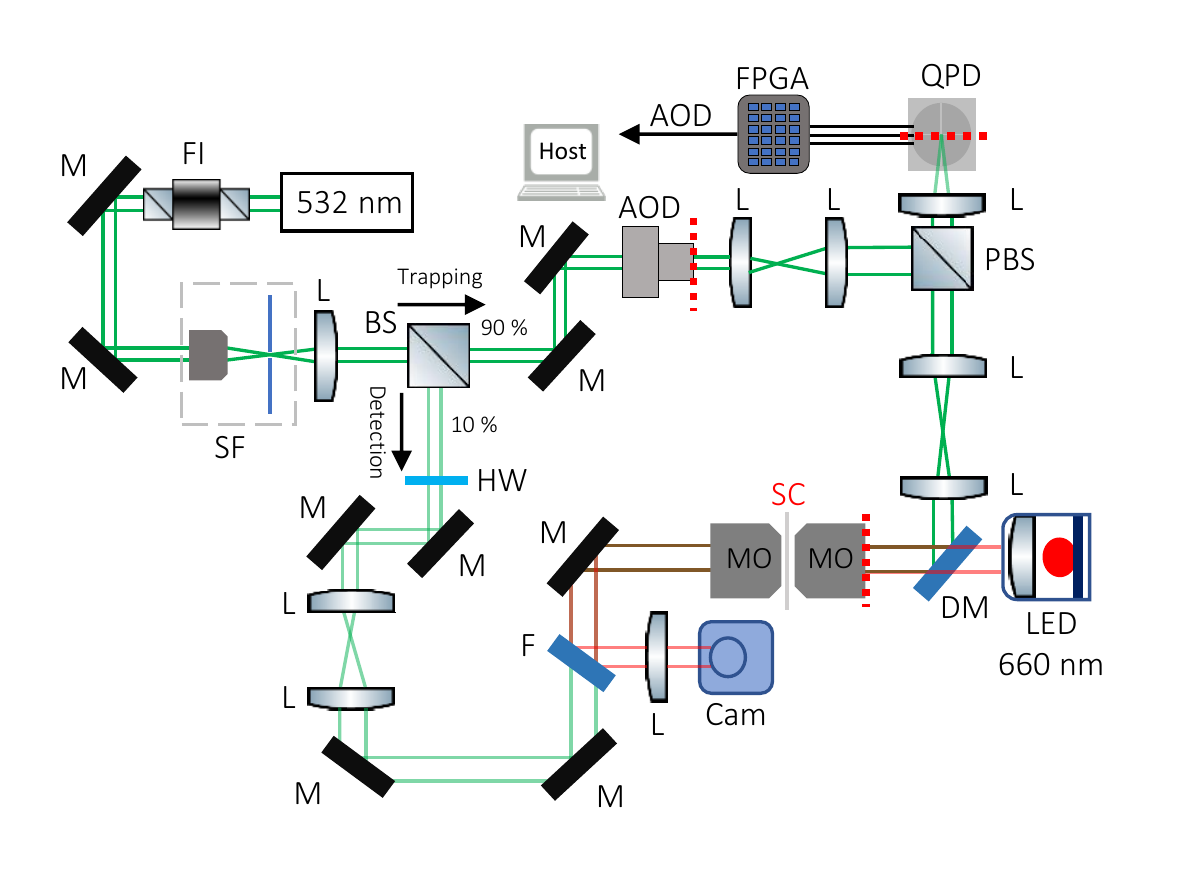}
\caption{\textbf{Schematic diagram of the feedback-trap setup.} FI = Faraday
isolator, M = mirror, SF = spatial filter, BS = beam splitter (non-polarizing), AOD = acousto-optic deflector, L = lens, MO = microscope
objective, SC = sample chamber, PBS = polarizing beam splitter, HW = half-wave plate, F = short-pass filter, QPD = quadrant photodiode, DM = dichroic mirror, PD = photodiode, Cam = camera. Planes conjugate to the backfocal plane of the trapping objective are shown in red-dashed lines.}
\label{fig:setup}
\end{figure}

A schematic diagram of the optical-tweezer-based feedback trap setup is shown in \ref{fig:setup}. The OT traps a colloidal particle (silica bead, \O 1.5 \textmu m, Bangs Laboratories) diffusing in water.   We use a 532-nm, solid-state laser (Nd:YAG, Coherent Genesis MX STM-series, 1 W) for trapping and detection. The polarization of the detection laser is rotated by 90$^{\circ}$ to  minimise interference with the trapping laser. The feedback forces originate from the shifting of the trap centre relative to the trapped particle position.  We use an acousto-optic deflector (DTSXY-250-532, AA Opto Electronic) to shift the trap centre, which is imaged at the back focal plane of the trapping objective to produce linear motion of the beam at the trapping plane.  The trapped particle scatters light that is collected in the forward direction by a microscope objective (Olympus UPLSAPO60XW water immersion, 60X, NA= 1.2).  A quadrant photodiode (First Sensor, QP50-6-18u-SD) is placed at the back focal plane of the objective to detect the particle motion. The signal from the photodiode is sent to a LabVIEW-based FPGA data acquisition (DAQ) system (NI 7855R). The DAQ receives the signal, calculates the voltage based on the required force, and generates it  every 10 \textmu s.

We construct a one-dimensional virtual tilted double-well potential using a feedback-optical tweezer\cite{kumar2018nanoscale} (\ref{fig:FB}). It is a continuous piecewise potential with a double well joined by linear potentials at the extremes. The overall potential is set in an asymmetric domain. The tilted double-well potential is parametrized as
\begin{equation}
	U_0(x) = E_\text{barrier} \left[ 1-2\left(\frac{x}{x_\text{m}} \right)^2
		+ \left( \frac{x}{x_\text{m}} \right)^4 \right] - \frac{1}{2}  
		\Delta E\left(\frac{x}{x_\text{m}} \right) \,,
\label{U}  
\end{equation}
where $E_\text{barrier} = 2$ is the barrier height, $\Delta E = 1.3$ the tilt in the potential, and $x_\text{m} = 40$ nm the well position. The energy $U_0(x)$ is scaled by $k_{\text{B}}T_\text{b}$ and length by $\sqrt{D\Delta t} \approx 1.8$ nm, where $k_{\text{B}}$ is the Boltzmann constant, $T_\text{b}$ the bath temperature (set by the room temperature), $D = 0.32$ \textmu m$^2$/s the diffusion coefficient of the particle, and $\Delta t = 10$ \textmu s the sampling time.  The bath temperature is typically $\approx 23~^\circ$C.  Its precise value for different runs is unimportant since the potential and all related energies are scaled by $k_\text{B}T_\text{b}$ and thus are independent of the bath temperature value.  We also note that the temperature quenches are large, never less than a factor of two in absolute temperature.  Minor temperature drifts during runs then do not directly have a significant effect on the dynamics of $p(x,t)$.

The overall potential energy landscape $U(x)$ of the bath is given as
\begin{equation}
	U(x) \equiv 
	\begin{cases}
 	  	U_0(x_\text{l}) + F_\text{max} x  & x \leq x_\text{l} \\[4pt]
		U_0(x) & x_\text{l} \leq x \leq x_\text{r} \\[4pt] 
		U_0(x_\text{r}) - F_\text{max} x  & x \geq x_\text{r}, \\[4pt]    
	\end{cases}
\label{eqn:pieceDW}
\end{equation}  
where $x_\text{l}$ and $x_\text{r}$ are positions defined so that $|U_0'(x_\text{l})| = |U_0'(x_\text{r})| = F_\text{max}$.  The potential $U(x)$ and its first derivatives are continuous everywhere, but the second derivative has  jump discontinuities at $x_\text{l}$ and $x_\text{r}$.  To implement the double-well potential in Eq.~\eqref{U} requires a force whose magnitude increases indefinitely at large distances from the well minima.  However, optical tweezers are limited to a maximum force, given a fixed beam power.  To accommodate this physical constraint, we match the double-well potential of Eq.~\eqref{U} beyond $x_\text{l}$ and $x_\text{r}$ to a linear potential whose slope corresponds to the maximum force an optical tweezer can exert.

\subsection{Choice of potential energy landscape \\ }

We have engineered our potential in such a way that the equilibration times for both hot and warm systems are $\lesssim 0.1$ s.  Such short times allow us to reach the equilibrium state with the bath, to connect directly to our definition of the Mpemba effect.  They also allow for easy acquisition of several thousand runs.  From such an ensemble, we can accurately reconstruct the time-dependent nonequilibrium state $p(x,t)$ of the system as it cools.  Moreover, because we recalibrate after each quench, we avoid the effects of drifts.  In particular, even after allowing all transient effects due to the preparation of an experiment to die away, we consistently observe drifts in position measurements on the order of  1 nm s$^{-1}$.  Given length scales of $\approx 100$ nm, these can become significant after several seconds.  By limiting runs to 0.1 s, we ensure that effects due to drifts are negligible.

Having chosen the overall scale of the potential, we needed to define its actual shape.  The barrier height and tilt are adjusted in such a way that the system, when trapped in the metastable state, takes longer to reach the equilibrium than the system that finds a direct path towards the equilibrium.  

As Eq.~\eqref{eqn:pieceDW} implies, we also impose a linear potential for $x<x_\text{l}$ and $x>x_\text{r}$.  The principal motivation, in our case, is that the tweezers can impose a maximum force $F_\text{max}$, and we simply allow the imposed force field to saturate when that limit is reached.  Because the maximum forces are large, we can reach large energies, the potential can easily range up to $\approx 100\, k_\text{B}T_\text{b}$. Such energy ranges are much larger than ordinary materials.  However, we use such large energy scales solely as a means to create short time scales.

\begin{figure}[t!]
\centering
\includegraphics[width=1\linewidth]{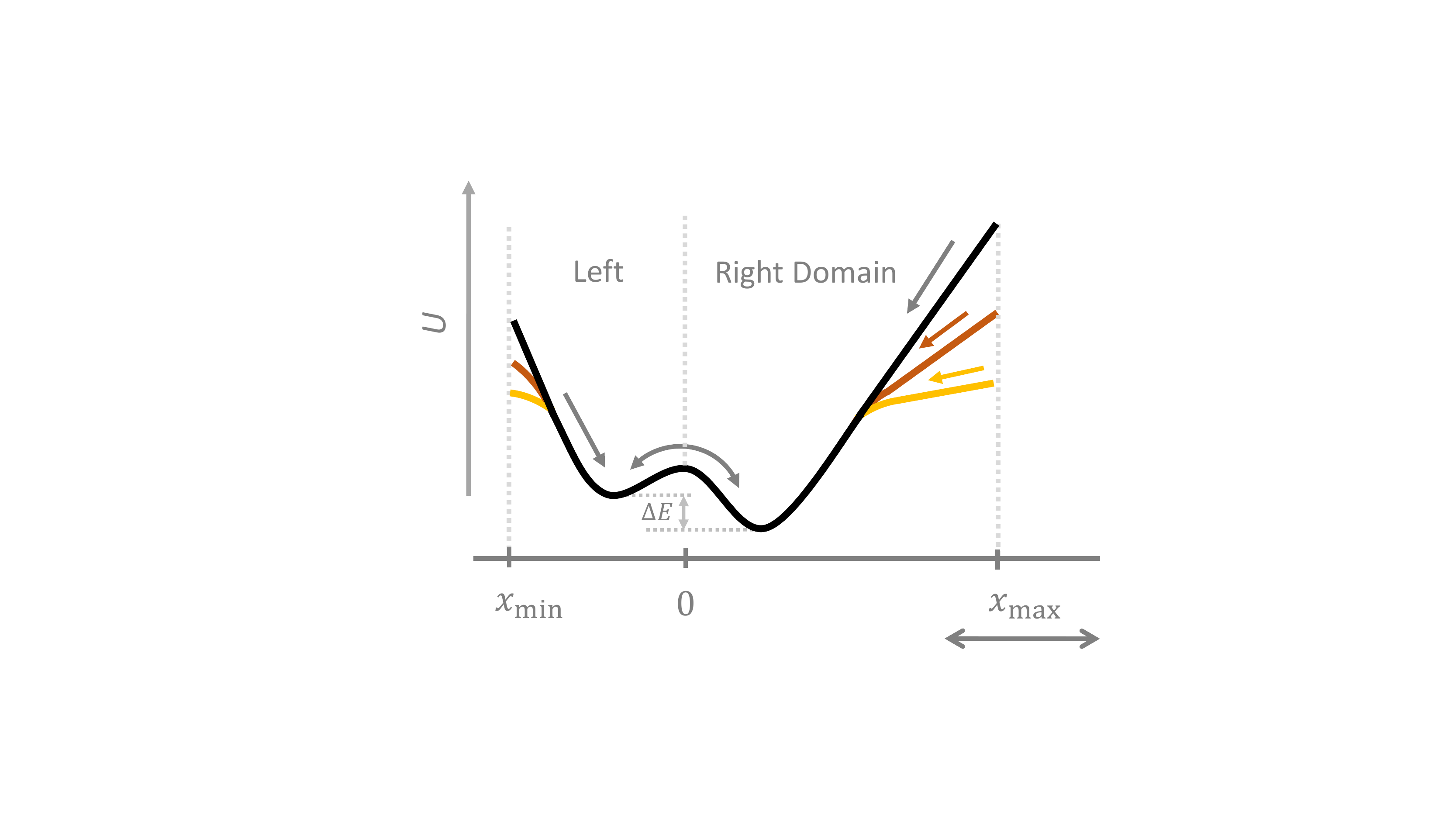}
\caption{\textbf{Potential energy landscape of the bath.} The bath potential energy is shown with different slopes for the kinetic path. A steep slope represents high velocities with which the particles are quenched towards the minima. The steepness of the linear potential determines both the time  and temperature scales.}
\label{fig:potential_choice}
\end{figure}

If time scales were allowed to be longer, then we could create similar dynamics with much reduced energy scales. The velocity at which the particle is pulled towards a minimum is determined by the force as $v_\text{max} = F_\text{max}/\gamma \approx E_\text{tot}/{\gamma\,\ell}\approx 60$ \textmu m/s, where $F_\text{max} \approx 20~\text{pN}$ ($0.2\, k_\text{B}T_\text{b}/\text{nm}$) is the maximum force exerted by the optical tweezer, $E_\text{tot} \approx 100$ the energy at the domain boundaries, $\gamma$ the viscous drag coefficient, and $\ell$ the distance between the basins of attraction and the respective domain boundaries. Thus, the kinetic timescale is approximated as $\tau \approx \ell^2/{D E_\text{tot}} \approx 0.3$ ms, where $D = k_\text{B}T_\text{b} / \gamma \approx 0.3$ \textmu m$^2$/s is the diffusion constant of the particle at the bath temperature $T_\text{b}$.  Thus, we choose a large energy scale to have a fast relaxation towards the two macrostates. If we were to use a lower maximum slope of potential, we would have the same overall structure and range of Mpemba effects, but their time scale would be correspondingly longer.  \ref{fig:potential_choice} illustrates qualitatively how similar effects can be seen in potentials where there is a reduced maximum slope.

\subsection{Accuracy of the imposed potential \\ }
\label{sec:accuracy}

We test directly our ability to impose a potential of a desired form, making use of the ``control'' data shown in Fig.~\ref{fig:Mp_Traj}.  The initial condition here is drawn from the same (nominal) distribution of the actual bath.  Our method is (1) impose an initial condition drawn from the Boltzmann distribution $\pi(x;T_\text{b}) \sim \exp [-U(x)/k_\text{B}T_\text{b}]$; (2) wait a time long compared to the equilibration time (Fig.~\ref{fig:Mp_Traj} shows that 60 ms suffices); (3) record the position.  We repeat the measurement $N=1000$ times, plot a position histogram, and infer the potential $U(x)/k_\text{B}T_\text{b}$ from the Boltzmann distribution $\pi(x;T_\text{b})$.  By using one data point from each run, we avoid issues due to correlated measurements.  Because we recalibrate positions after each run, we minimise the effects due to drift.  Finally, this \textit{ensemble method} matches the one used in the cooling experiment.

\begin{figure}[h!]
\centering
\includegraphics{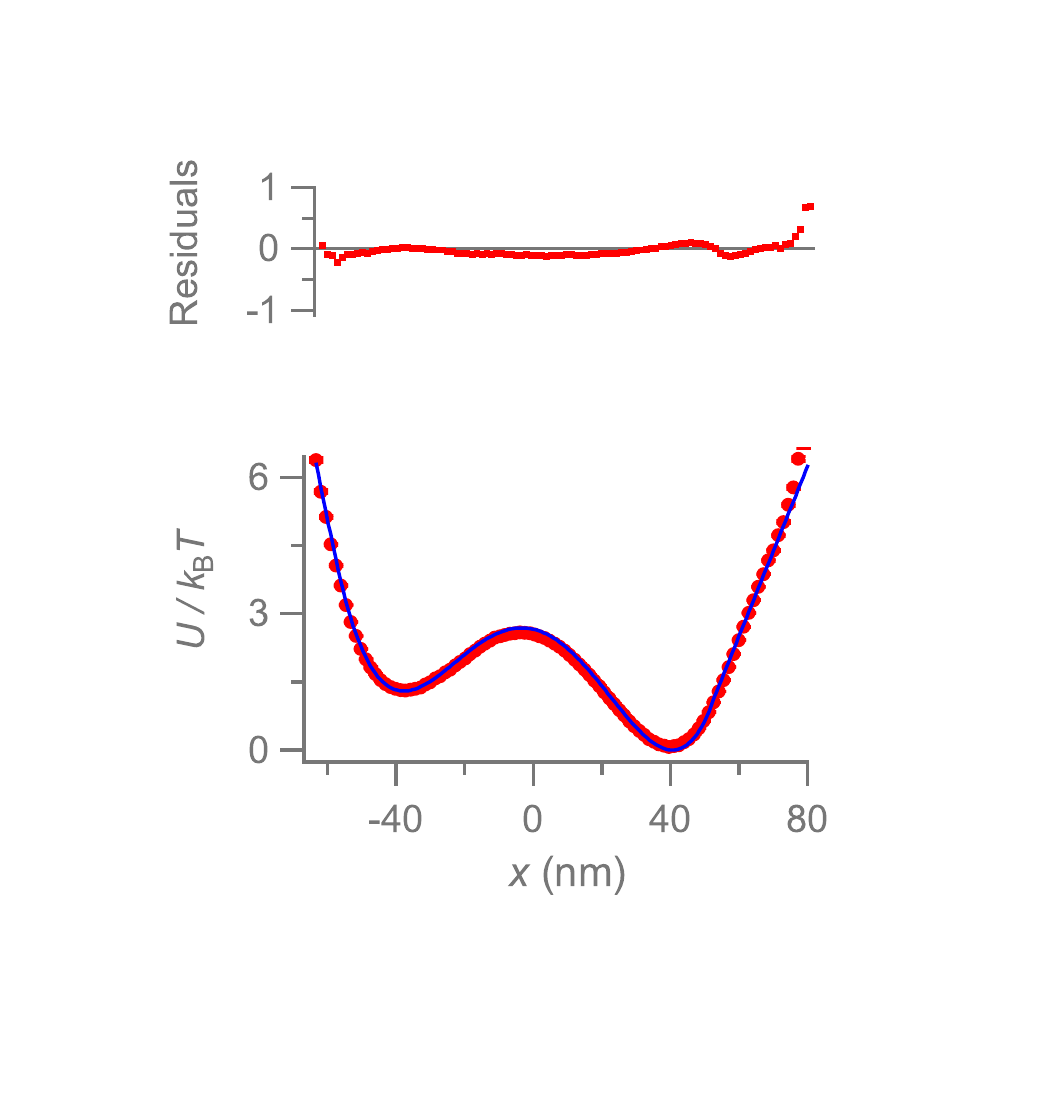}
\caption{\textbf{Potential energy landscape of the bath.} Red markers denote the potential
reconstructed from the Boltzmann distribution of the position measurements, with no curve fitting; the superimposed solid black line shows the imposed potentials.  The error bars represent $\sqrt{N_\text{b}}/N$, where $N_\text{b}$ is the number of counts in each bin and $N$ the total number of counts.}
\label{fig:BoltzmannTest}
\end{figure}

\ref{fig:BoltzmannTest} shows the reconstructed potential from the position measurements. The RMS error of the residuals is 0.18 $k_\text{B}T_\text{b}$. Because of the limited statistics possible with the direct Boltzmann measurement, we restrict the reconstruction to a range of energies $6~k_\text{B}T_\text{b}$.

\section*{Infinite potential vs.~finite potential} 
\label{sec: FinVsInf}

\begin{figure}[ht!]
\centering
\includegraphics{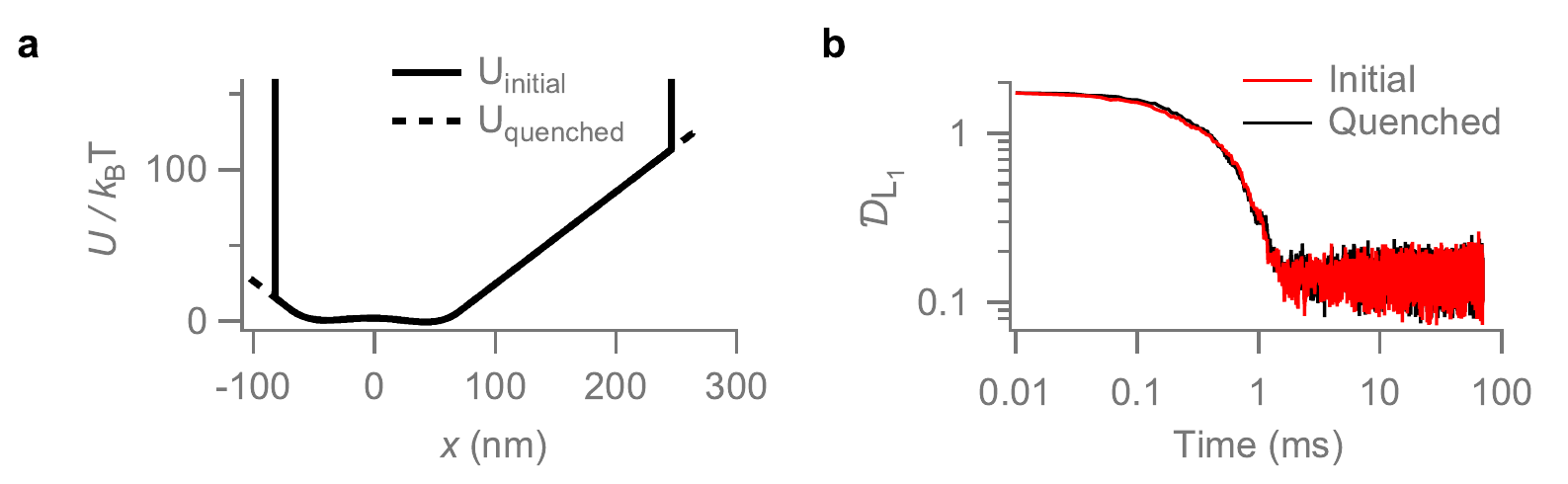}
\caption[Pot_fin_inf] {
\textbf{Finite maximum slope of the potential does not affect particle dynamics significantly.} \textbf{a}, The energy landscape for the Mpemba effect. Solid line depicts the initial energy landscape with infinite potential walls at the domain boundaries. The equilibrium distribution of the particle is calculated based on this potential ($U_{\text{initial}}$).  Dashed line shows the potential ($U_{\text{quenched}}$) in which the particle is quenched. \textbf{b}, Langevin simulations of the Mpemba effect using both potentials show no significant differences between the two cases.}
 \label{fig:Pot_fin_inf} 
 \end{figure}

Another systematic deviation in the imposed potential from the desired shape arises because the initial conditions for the cooling experiment were calculated assuming infinite walls at the domain boundaries.  However, physically imposing an infinite potential wall is impossible, meaning that there is a maximum possible force exerted by the virtual potential while the particle is evolving towards equilibrium with the bath. Nevertheless, we can and do take into account the infinite walls in creating the initial conditions for particles released in the potential.  In almost all cases, the inward forces cause the particles to move towards positions in the interior of the experimental domain.  In rare cases, a fluctuation from the bath can briefly push a particle outside the domain defined by the infinite walls.  Thus, particles moving in the physically imposed potential can have motion that violates very slightly the potential assumed in defining the initial conditions and assumed in calculating quantities such as the eigenfunctions of the FPE.

To test whether such violations are important, we simulate the overdamped particle motion in the feedback trap using a discretised Langevin equation\cite{jun12}
\begin{align}
x_\text{n+1} &= x_\text{n}+\frac{1}{\gamma}F_\text{n}\,\Delta t+\xi_\text{n}\,, \nonumber \\[4pt]
\bar{x}_\text{n+1} &= x_\text{n}+\chi_\text{n}\,,
\label{Eq: Langevin}
\end{align}
where $x_\text{n}$ is the true position of the particle, $\bar{x}_\text{n}$ the observed position at time $t_\text{n}$, and $\xi_\text{n}$ and $\chi_\text{n}$ are the integrated thermal and measurement noises. The force $F_\text{n} = - \partial_x\,{U}(\bar{x}_\text{n},\Delta t)$ is applied at a deterministic time step of $\Delta t = 10$ \textmu s. 

Langevin simulations based on Eq.~\eqref{Eq: Langevin} for both the idealized and physical potentials (\ref{fig:Pot_fin_inf}) show that these small violations have no significant effect on the quantity of interest, the distance function $\mathcal{D}$.

\section*{Imposing an instantaneous quench via initial conditions} 
\label{sec:quench}

\begin{figure}[t!]
\centering
\includegraphics[width=0.6\linewidth]{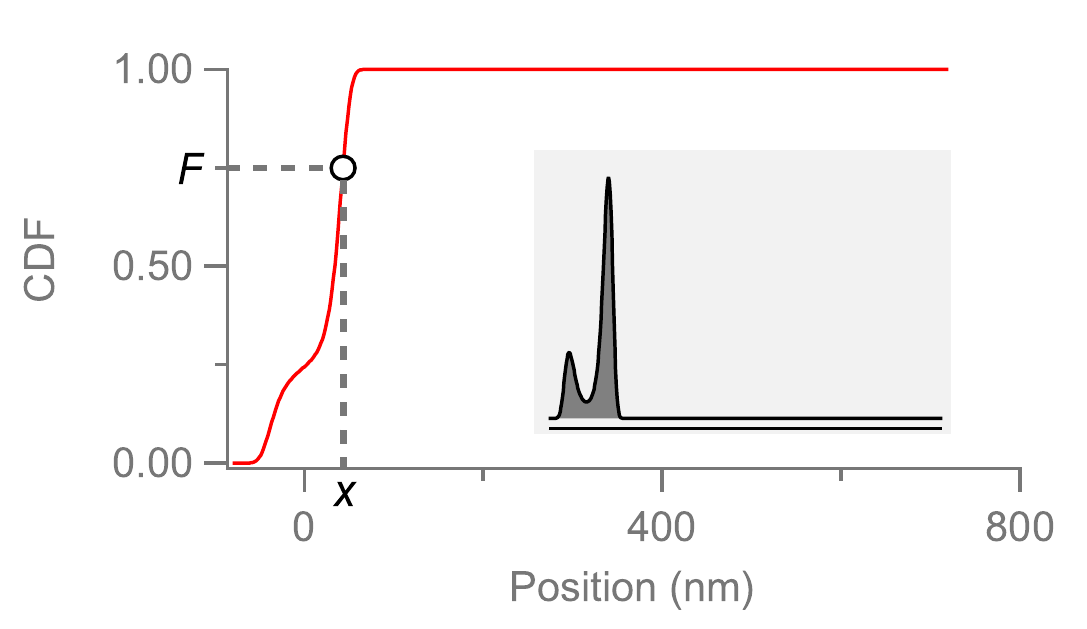}
\caption{\textbf{Cumulative probability distribution at the bath temperature.}  The cumulative distribution of the Boltzmann distribution (inset) at the bath temperature is calculated.  An algorithm based on binary search and linear interpolation is used to map the CDF ($F$) to position $x$ (dashed lines). Asymmetry coefficient $\alpha = 9$.}
\label{fig:cdf}
\end{figure}

The initial probability distributions correspond to Boltzmann distributions at $T_\text{initial}$.  However, physically preparing systems that are in thermal equilibrium at high temperatures such as $T_\text{initial} = 1000\,T_\text{b}$ is not possible in our setup. Nor is it possible to create an instantaneous quench by changing the temperature of the bath. Instead, we sample initial positions from an equilibrium distribution and place the particle at those positions in the beginning of each run. To implement this, we calculate the  cumulative distribution function (CDF) from the equilibrium probability density function (PDF)\cite{press07,zhang19}. The CDF for a random variable $X$ is given as
\begin{align}
	F_{X}(x) = \int_{x_\text{min}} ^{x} \dd{x} p(x) \,,
\end{align}
where the PDF $p(x)$ is integrated over the range $[x_\text{min},x]$ to calculate the CDF, $F_{X}(x)$.  Since the CDF is in the range $[0,1]$, we use a uniform random number generator to generate numbers between 0 and 1. Initially a binary-search algorithm is used to find the corresponding position.  If the random number is not found by the binary search, a linear interpolation based on the neighbouring values is used to get the accurate position (\ref{fig:cdf}).  We create lookup tables (LUTs) of the CDF functions at different temperatures and sample the initial position in a similar way for each run. Because the initial potential includes hard walls, the probability to draw an initial condition with $x < x_\text{min}$ or $x>x_\text{max}$ is zero.  We thus normalize the PDF and CDF on the range $(x_\text{min}, x_\text{max})$.

\section*{Measuring the distance to equilibrium} 
\label{sec:DF}

Consider a colloidal particle immersed in a fluid bath of temperature $T_\text{b}$ and subject to a one-dimensional potential $U(x)$.  For systems in thermal equilibrium, the position $x$ of the particle, when sampled from an ensemble of identically prepared systems, will obey the Boltzmann distribution
\begin{align}
	\pi(x; T_\text{b}) = \frac{1}{Z} \e^{-U(x)/k_\text{B}T_\text{b}} \,,
\label{eq:boltzmanndist}
\end{align}
where $k_\text{B}$ is Boltzmann's constant and where the partition coefficient $Z=\int \dd{x} \exp [-U(x) / k_\text{B}T_\text{b}]$ normalizes the probability distribution.

For a nonequilibrium system, it is not possible, in general, to define an equivalent notion of temperature.   In a macroscopic system such as the ones used for previous experiments on the Mpemba effect, the system is typically in local equilibrium and may be described by a temperature field, $T(\bm{x},t)$.  When subject to the temperature quench specified by the protocol used in the Mpemba effect, temperature gradients are large, and it is impossible to characterize the system accurately by a single time-dependent temperature, such as the spatial average of $T(\bm{x},t$). In addition, a fluid object can have internal fluid motions that arise because of the quench (such as convection currents created when the top cools off more quickly than the bottom), meaning that other fields may be relevant, too.

In the mesoscopic single-colloidal-particle system studied here, we can measure the instantaneous probability distribution $p(x,t)$, the probability density for the measured position to lie between $x$ and $x+\dd{x}$, by conducting a series of experiments on identically prepared trials.  The set of trials forms an ensemble.

Lu and Raz\cite{lu2017nonequilibrium} have argued that the observation of the Mpemba effect is independent of the choice of the functional that measures the distance from thermal equilibrium if the measure satisfies three properties:
\begin{enumerate}
\item $\mathcal{D}[p(x,t), \pi(x,T_\text{b})]$ should be a monotonically non-increasing function of time during relaxation towards equilibrium;

\item $\mathcal{D}[\pi(x,T_\text{initial}),\pi(x,T_\text{b})]$ should be a monotonically increasing function of $T_\text{initial}$ for all\\ $T_\text{initial}~>~T_\text{b}$, so that initially hotter states are farther from the bath distribution;

\item $\mathcal{D}[p(x,t))]$ should be a continuous and convex function of probability $p$ when evaluated at any particular value of $x$ and $t$.
\end{enumerate}
Although we often write $\mathcal{D}$ using a simplified notation that omits terms from its arguments, it is important to remember that it is a functional that depends on both a dynamic probability distribution $p(x,t)$ and a reference equilibrium distribution $\pi(x, T_\text{b})$.  Note that the measure is not required to be a proper distance, allowing the asymmetric Kullback-Leibler (KL) divergence as one possibility.

The results shown in Fig.~\ref{fig:Mp_temp_alpha} (main text) are based on $L_1$ distance measures.  We begin by defining and discussing the $L_1$ distance measure here.  We then check that  similar results are found using the KL divergence.  In the main text, we use the notation $\mathcal{D}$ (with no subscript) to represent the $L_1$ distance; here, we will use a subscript to differentiate between different distance measures.

\subsection{$\bm{L_1}$ distance.}

To evaluate this distance from trajectory data, we partition the position measurements into $N_\text{b}$ bins:
\begin{equation}
	\mathcal{D}_{L_1}[p(x,t);\pi(x;T_\text{b})] = \sum_{i=1}^{N_\text{b}}{|p_i-\pi_i|} \,,
\label{Eq:dist_L1}
\end{equation} 
where $p_i \equiv p(x_i,t)$ is the frequency estimate of the probability for a measured position $x$ at a time $t$ after the quench to fall within the interval $[x_i, x_{i+1})$, where $x_i \equiv i \, \Delta x$, with $\Delta x = ( |x_\text{max}| + |x_\text{min}|) / N_\text{b}$.  Similarly, $\pi_i \equiv \pi(x_i;\,T_\text{b})$ is the histogram estimate of the Boltzmann distribution  at temperature $T_\text{b}$. The smallest $L_1$ distance measured between the two distributions is limited by the statistical noise due to the finite sample size.  To make a naive calculation for two uniform distributions, we can write
\begin{align}
	p_i = \frac{N_\text{c}}{N}\,,
\end{align}
\noindent where $N_\text{c} = N/N_\text{b}$ is the average number of counts in each bin and $N$ the number of trials. The variance, $\sigma^2$ of $p_i$ for a typical bin is approximately
\begin{align}
	\sigma_{p_i}^2 = \frac{N_\text{c}}{N^2} = \frac{1}{N_\text{b}N} \,.
\end{align}
The variance of $|p_i - \pi_i|$ is expected to be comparable.  Then, summing over $N_\text{b}$ bins and taking a square root to estimate the standard deviation leads us to expect fluctuations of
\begin{align}
	\sigma_{\mathcal{D}_{L_1}}  = \mathcal{O} \left( \sqrt{\frac{N_\text{b}}{{N}}} \right) \,.
\label{Eq: L1error}
\end{align}
Numerically, we confirm this scaling of fluctuations.  A more sophisticated approach---not needed here---would be to calculate the mean absolute difference of two Poisson variables, which can be expressed in terms of a Skellam distribution.  The main point is that the noise level scales with the number of trials $N$ that constitute the ensemble as $N^{-1/2}$.

\subsection{Kullback-Leibler (KL) divergence.}

Using similar definitions of the distributions, we write
\begin{align}
	\mathcal{D}_\text{KL}\left[ p(x,t);\, \pi(x;T_\text{b}) \right] 
	\equiv \sum_{i = 1}^{ N_\text{b}}p_i \ln \left( \frac{p_i}{\pi_i}\right) 
	= \sum_{i = 1}^{ N_\text{b}} p_i \ln p_i - p_i \ln \pi_i \,.
\label{eq:KL}
\end{align}
\noindent The KL divergence measures the relative entropy between two probability distributions.  It has a physical motivation in that it can be related to the nonequilibrium free energy of a system as $F_\text{noneq} = F_\text{eq} + \mathcal{D}_\text{KL}\left[ p(x,t);\, \pi(x;T_\text{b}) \right]$ and can thereby be connected to the entropy produced during the relaxation process\cite{lu2017nonequilibrium}. Both equilibrium and nonequilibrium free energies here are scaled by  $k_\text{B}T_\text{b}$.

\begin{figure}[ht!]
\centering
\includegraphics[width=3.0in]{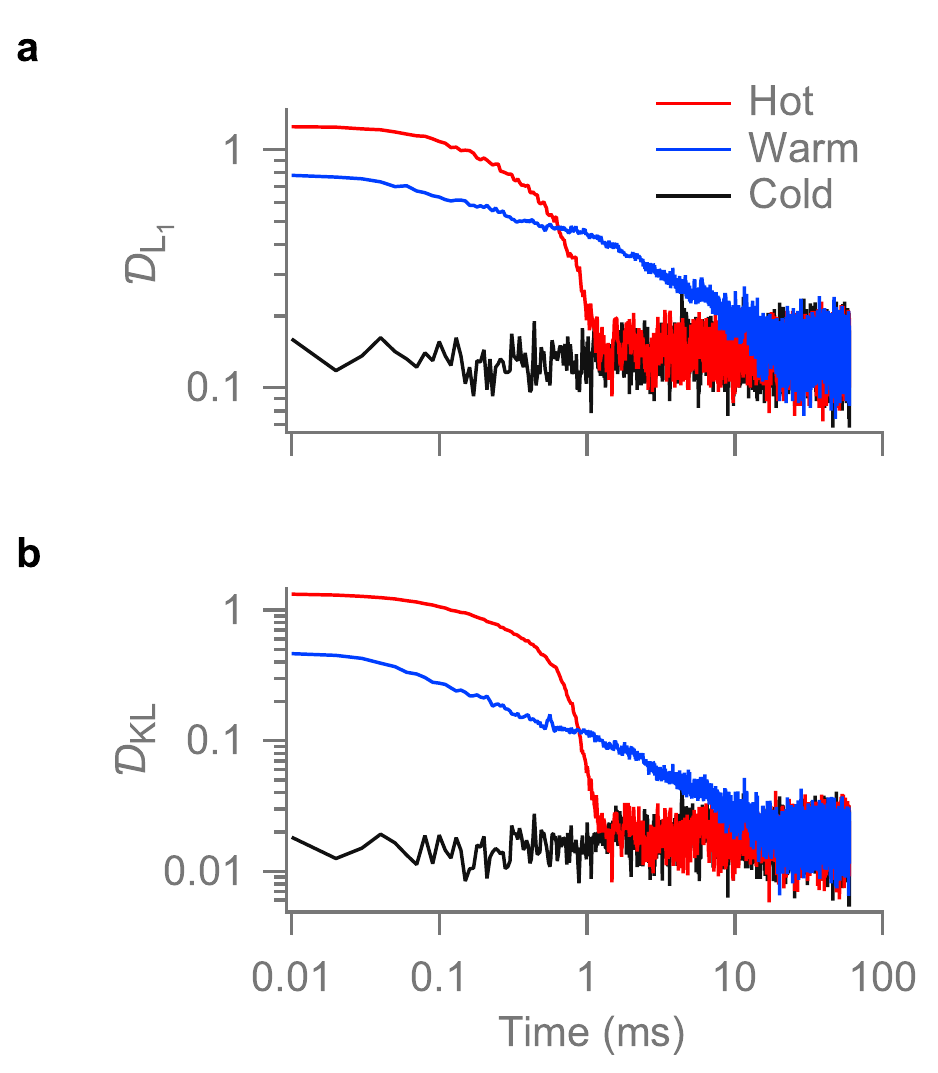}
\caption{\textbf{Mpemba effect is robust to choice of distance measure.}  \textbf{a} and \textbf{b} show respectively the $L_1$ and Kullback-Leibler distances.  Data corresponds to Fig.~\ref{fig:Mp_Traj} (main text) for hot ($T_\text{h} = 1000$), warm ($T_\text{w} = 12$), and cold ($T_\text{c} = 1$) temperatures.  Both distance measures show crossing, indicating the Mpemba effect.}
\label{fig:dist_allMP}
\end{figure}

In \ref{fig:dist_allMP}, we show both distance measures for the Mpemba effect based on the data presented in Fig.~\ref{fig:Mp_Traj} of the main text.  The figure illustrates that the observation of the Mpemba effect does not depend on the choice of distance function.  The numerical details and shape of the individual curves may change, but the crossing of curves is a robust observation. 

Although the KL divergence gives qualitatively similar results, it has two inconvenient features that lead us to prefer the $L_1$ distance.  The first is that some bins will have zero counts.  If these zero-value bins were counted in Eq.~\eqref{eq:KL}, the measured KL divergence would be infinite. To avoid such issues, we regularise the equilibrium distribution by adding a single pseudocount to each bin\cite{jaynes2003probability}.  We then normalize the histogram to estimate the probability density.  Although the use of pseudocounts biases the distance estimation slightly, there is no effect on the presence or absence of distance-curve crossing, as demonstrated by the fact that the results with the KL divergence agree qualitatively with those using the other distance measures.  

The second inconvenience of the KL divergence is that to extract the $a_2$ coefficient requires a Taylor expansion, which is not needed when using the $L_1$ norm.

% the relation between $\mathcal{D}_\text{KL}$ and the $a_2$ coefficient is more involved, making it less convenient a tool for connecting between empirical data and theoretical predictions based on eigenfunction analysis for the $a_2$ coefficient.

\section*{Eigenfunction analysis}
\label{sec:eigenfunc}

In our experiment, the particle is continuously under the influence of drag forces and random forces.  The time evolution of the particle is generally described by the Langevin equation (as used, in discretised form, in Methods, Eq.~\ref{Eq: Langevin}). Equivalently, these dynamics can also be described in terms of the time evolution of the probability density $p(x,t)$ of particle positions by the Fokker-Planck (FP) equation as
\begin{align}
	\frac{\partial{p}(x,t)}{\partial{t}} = \left[-\frac{1}{\gamma}\frac{\partial{}}{\partial{x}}F(x)
		+ \, \frac{k_\text{B}T_\text{b}}{\gamma} \frac{\partial{}^2}{\partial{x^2}}\right]\,p(x,t) \equiv \mathcal{L}\,p(x,t) \,,
\label{eq:FP}
\end{align}
where $\mathcal{L}$ is the Fokker-Planck operator for the Brownian motion with $F = -\partial_x U(x)$. For heavily overdamped dynamics, the velocity variables that would otherwise be present in the FP equation may be neglected.  In this limit, the FP equation is sometimes referred to as the Smoluchowski equation\cite{risken89}. The solution $p(x,t)$ of the FP equation in terms of its eigenfunctions is given as
\begin{align}
	 p(x,t) = \pi(x;T_\text{b}) + 
		\sum _{k = 2}^{\infty}{a_k(\alpha,T_\text{initial}) \e^{-\lambda_k t}v_2(x; \alpha,T_\text{b})} \,,
\label{Eq:FPsolution}
\end{align}
where $\pi(x;T_\text{b})$ is the equilibrium probability density function, achieved for $t\to \infty$, $v_2(x; \alpha,T_\text{b})$ and $\lambda_k$ the $k^\text{th}$ right eigenfunction / eigenvalue pair (assumed non-degenerate), which are ordered such that $0 = \lambda_1 < \lambda_2 < \cdots$. For $\lambda_2 < \lambda_3$, the higher-order terms are exponentially small; thus, the eigenvalue $\lambda_2$ corresponds to the slowest relaxation rate. Note that relaxation $\sim e^{-\lambda_3 t}$ is exponentially faster than relaxation $\sim e^{-\lambda_2 t}$, so that the condition $a_2=0$ corresponds to an exponential speed-up of relaxation rate. At very long times, Eq.~\eqref{Eq:FPsolution} implies that $ p(x,t) \approx \pi(x;T_\text{b})$, meaning that initial conditions corresponding to any given temperature eventually all relax to the same equilibrium state, with temperature $T_\text{b}$.

\subsection{Adjoint of the Fokker-Planck operator.}
\label{sec:AdjointFP}

In our system, the probability density function for a particle to be found at position $x$ at a time $t$ after a quench, $p(x,t)$, obeys the Fokker-Planck (FP) equation $\partial_t p(x,t) = \mathcal{L}\, p(x,t)$.

In Eq.~\eqref{eq:FP}, the force is $F(x) = -\partial_x U(x)$, and the probability density function $p(x,t)$ obeys the boundary condition
\begin{align}
	\mathcal{J}(x_\text{min}) = \mathcal{J}(x_\text{max}) = 0 \,,
\end{align}
where the probability current $\mathcal{J}(x)$ is defined to be
\begin{align}
	\mathcal{J}(x) \equiv \frac{F}{\gamma}(x)\,p(x) -  \frac{k_\text{B}T_\text{b}}{\gamma} \pdv{p}{x} \,.
\end{align}
Physically, the boundary condition expresses the fact that the particle is in thermal equilibrium (no current) everywhere, including at the boundaries.

As we will see below, in order to apply the analysis developed by Lu and Raz\cite{lu2017nonequilibrium}, we need to evaluate (numerically) not only the right eigenfunction $v_2(x; \alpha,T_\text{b})$ but also the associated left eigenfunction $u_2(x; \alpha,T_\text{b})$ of the adjoint $\mathcal{L}^{\dagger}$ of the FP operator\cite{stone2009mathematics}.  One subtlety is that the boundary condition for $\mathcal{L}^{\dagger}$ differs from that of the $\mathcal{L}$ operator. To find $\mathcal{L}^{\dagger}$ and its boundary conditions, we introduce two test functions $\phi(x)$ and $p(x)$ and evaluate the inner product
\begin{align}
	\left\langle\phi | \mathcal{L}\, p\right\rangle 
	&= -\int_{x_\text{min}}^{x_\text{max}} \dd{x}\, \left[ \phi(x) \, \pdv{x} \, 
		\left[ \frac{F(x)}{\gamma} p(x) \right] \right]
	+ \frac{k_\text{B}T_\text{b}}{\gamma} \int_{x_\text{min}}^{x_\text{max}} \dd{x} \, \phi(x)\left( \pdv[2]{p}{x} \right)
\end{align}
Evaluating both integrals by parts, we can write
\begin{align}
	\left\langle\phi | \mathcal{L}\, p\right\rangle 
	&= \int_{x_\text{min}}^{x_\text{max}} \dd{x} \, \left.\left[ \frac{F(x)}{\gamma} \left( \pdv{\phi}{x} \, 
		 \right) p(x) \right] - \left[ \phi(x) \frac{F(x)}{\gamma} p(x) \right] \right|^{x_\text{max}}_{x_\text{min}}			\nonumber\\
	&\qquad -  \frac{k_\text{B}T_\text{b}}{\gamma} \int_{x_\text{min}}^{x_\text{max}} \dd{x} \,
		\left( \pdv{\phi}{x} \, \pdv{p}{x} \right)
		+  \frac{k_\text{B}T_\text{b}}{\gamma} \left[ \phi(x) \pdv{p}{x} \right]^{x_\text{max}}_{x_\text{min}} \,.
\label{eq:adjointderivation0}
\end{align}
Integrating the $k_\text{B}T_\text{b}/\gamma$ term again by parts gives
\begin{align}
	\left\langle\phi | \mathcal{L}\, p\right\rangle 
	&= \int_{x_\text{min}}^{x_\text{max}} \dd{x} \, \left[ \frac{F(x)}{\gamma} \left( \pdv{\phi}{x} \right) 
		+ \frac{k_\text{B}T_\text{b}}{\gamma} \pdv[2]{\phi}{x} \right] p(x) \, \nonumber \\
	&\qquad
		+ \left[ - \phi(x) \frac{F(x)}{\gamma} p(x) + \phi(x)  \frac{k_\text{B}T_\text{b}}{\gamma} \, \pdv{p}{x} 
		-  \frac{k_\text{B}T_\text{b}}{\gamma} \left( \pdv{\phi}{x} \right) p(x) 
			\right]^{x_\text{max}}_{x_\text{min}}   \nonumber \\
	&= \left\langle\mathcal{L}^{\dagger}\phi | p\right\rangle-\, \left[ \phi(x) \, \mathcal{J}(x) + 
		 \frac{k_\text{B}T_\text{b}}{\gamma} \left( \pdv{\phi}{x} \right) p(x) \right]^{x_\text{max}}_{x_\text{min}} \,.
\label{eq:adjointderivation}
\end{align}
 Thus, the adjoint operator is $\mathcal{L}^{\dagger} = \frac{F(x)}{\gamma} \partial_{x} + \frac{k_\text{B}T_\text{b}}{\gamma} \partial_{xx}$.  Since $\mathcal{L} \neq \mathcal{L}^{\dagger}$, the FP operator is not Hermitian.  From Eq.~\eqref{eq:adjointderivation}, we see that $\mathcal{L}^{\dagger}$ obeys Neumann boundary conditions, 
\begin{align}
	\left. \pdv{\phi}{x} \right|_{x = x_\text{min}} = 
	\left. \pdv{\phi}{x} \right|_{x = x_\text{max}} = 0 \,,
\end{align}
in contrast to the boundary condition of zero probability flux, $\mathcal{J}(x_\text{min}) = \mathcal{J}(x_\text{max}) = 0$, obeyed by $\mathcal{L}$.

\subsection{Calculation of the \texorpdfstring{$a_2$}{a2} coefficient.}
\label{sec:a2}

\begin{figure}[ht!]
\centering
 \includegraphics{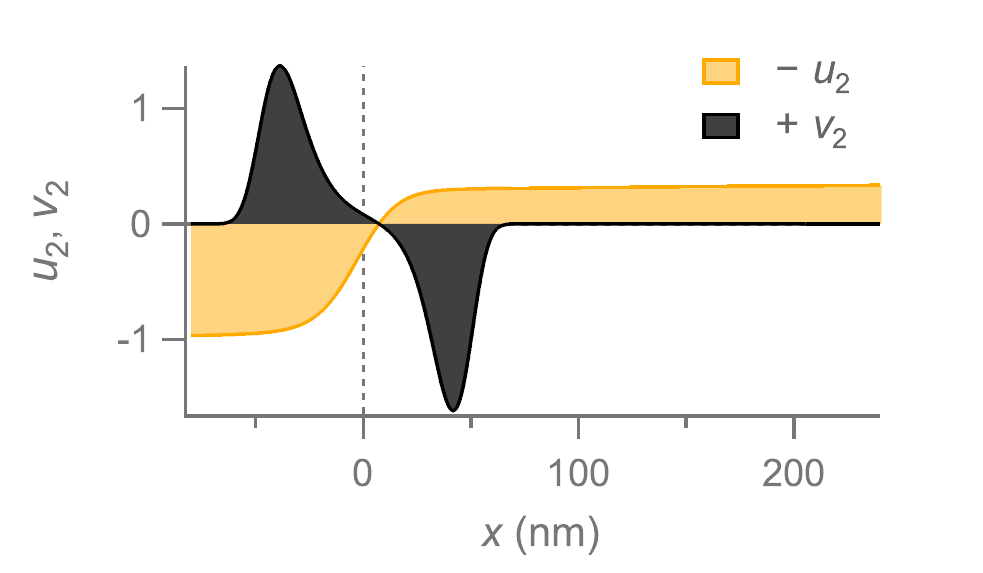}
\caption {\label{fig:eigenfunctions} 
\textbf{Eigenfunctions of the Fokker-Planck operator.} $\text{u}_2(x)$ and $\text{v}_2(x)$ are the left and right eigenfunctions, respectively, and correspond to the smallest non-zero eigenvalue of the FP operator. The negative of the left eigenfunction is plotted to aid to better visualisation. The eigenfunctions are calculated for $\alpha = 3$.}
\end{figure}

In our experiments, a particle is initially in equilibrium at a temperature $T_\text{initial}$ and then released into a bath at temperature $T_\text{b}$.  We will take the bath temperature as fixed but consider various initial temperatures $T_\text{initial}$.  Another relevant parameter is the asymmetry parameter $\alpha$ of domain sizes. The solution to the FP equation in terms of its eigenfunctions is then given as
\begin{align}
	p(x,t) = \pi(x;T_\text{b}) + \sum _{k = 2}^{\infty}{a_k(\alpha,T_\text{initial}) 
		\e^{-\lambda_k t}v_k(x; \alpha,T_\text{b})} \,,
\label{p1}
\end{align}
where $\{v_k (x) \}$ are the right eigenfunctions and $\{\lambda_k\}$ the corresponding eigenvalues of the FP operator.  At time $t=0$, the probability density is a Boltzmann distribution at the initial temperature $T_\text{initial}$:
\begin{align}
	 p(x,0) = \pi(x;T_\text{initial}) = \pi(x;T_\text{b}) + \sum _{k = 2}^{\infty}{a_k(\alpha,T_\text{initial}) v_k(x; \alpha,T_\text{b})} \,. 
\label{eq:p_initial}
\end{align}
We have shown that the FP operator is non-Hermitian, and thus, the left and right eigenfunctions are different. We numerically solve the FP equation for our system using standard \textit{Mathematica} operations to find the eigenfunctions.  \ref{fig:eigenfunctions} shows the negative left and positive right eigenfunctions corresponding to the smallest non-zero eigenvalue of $\mathcal{L}$.  To clearly show both the eigenfunctions, we have plotted the \textit{negative} left eigenfunction here.  

To calculate $a_2(\alpha,T_\text{initial})$, we evaluate the inner product between $u_2(x; \alpha,T_\text{b})$ and the initial probability distribution $p(x,0) = \pi(x;T_\text{initial})$.  Then, 

\begin{align}
	\langle u_2  | \pi(x;T_\text{initial}) \rangle = \langle u_2 | \pi(x;T_\text{b})\rangle 
	+ \sum _{k = 2}^{\infty}{a_k(\alpha,T_\text{initial}) \langle u_2 | v_k \rangle} \,,
\label{eq:u2projection}
\end{align}
where the inner product between two functions $f(x)$ and $g(x)$ in the interval $\left[ x_\text{min}, x_\text{max} \right]$ is defined as $\langle f | g\rangle \equiv \int^{x_\text{max}}_{x_\text{min}} \dd{x} \, f(x)g(x)$. Since the left and right eigenfunctions are biorthogonal, $\langle u_2 | v_k \rangle = 0$ for $k \neq 2$, and we evaluate the scalar products in Eq.~\eqref{eq:u2projection} to find $\langle u_2 | \pi(x;T_\text{initial}) \rangle = {a_2(\alpha,T_\text{initial})\langle u_2 | v_2 \rangle}$, or
\begin{align}
	a_2(\alpha,T_\text{initial}) 
		= \frac{\langle u_2 | \pi(x;T_\text{initial}) \rangle}{\langle u_2 | v_2 \rangle} \,,
\label{eq:a2}
\end{align}
where the normalization factor $\langle u_2 | v_2 \rangle = 0.55$, given our normalization convention, which is to take $\langle u_i | u_i \rangle = \langle v_i | v_i \rangle = 1$, for $i=1, 2, \cdots$.  In Eq.~\eqref{eq:a2}, we recall that $u_2(x)$ and $v_2(x)$ depend on the \textit{bath} temperature, $T_\text{b}$ and the asymmetry coefficient, $\alpha$.

\subsection{Relationship between \texorpdfstring{$\Delta \mathcal{D}$}{D} and the \texorpdfstring{$a_2$}{a2} coefficient} 
\label{sec: Delta_D_vs_a2}

In the experiment, we do not measure the second left and right eigenfunctions directly but rather a quantity $\Delta \mathcal{D}$ proportional to $|a_2(\alpha,\,T_\text{initial})|$.  To connect these quantities in the long-time limit, we rearrange Eq.~\eqref{p1} as $p(x,t) - \pi(x; T_\text{b})\,\approx\, a_2(\alpha,\,T_\text{initial}) \e^{-\lambda_2 t} v_2(x)$ for $k = 2$ and take the absolute difference between the dynamic and reference probabilities. However, in the experiment, we calculate the frequency estimate of the probability and thus, summing over all the bins for the absolute difference between the two probabilities gives
\begin{align}
	\sum^{N_\text{b}}_{i = 1}{| p_i - \pi_i|} 
	= \sum^{N_\text{b}}_{i = 1}|a_2(\alpha,\,T_\text{initial}) \e^{-\lambda_2 t} v_i|  + \sigma_\mathcal{D}\,,
\label{Eq: L1}
\end{align}
where $v_i \equiv v_2(i \Delta x)$ for $x \in [i\Delta x,(i+1)\Delta x)$ and $\sigma_\mathcal{D}$ is the noise in the $\mathcal{D}$ measurement due to finite sampling (Eq.~\ref{Eq: L1error}). The left-hand term in Eq. \ref{Eq: L1} is the $L_1$ distance between  the discretized distributions $p(x,t)$ and $\pi(x; T_\text{b})$.  Thus,
\begin{align}
	\mathcal{D}[p(x,t);\,\pi(x;T_\text{b})] 
		\equiv \mathcal{D}(t) &= |a_2(\alpha,\,T_\text{initial})| \e^{-\lambda_2 t}\sum^{N_\text{b}}_{i = 1}|v_i| + \sigma_\mathcal{D} \,,\nonumber \\
		&= |a_2(\alpha,\,T_\text{initial})| \e^{-\lambda_2 t} V + \sigma_\mathcal{D}\,,
\label{Eq:a2vsdelD}
\end{align}
where $V \equiv \sum^{N_\text{b}}_{i = 1}|v_i|$. The $\mathcal{D}(t)$ plot typically has two regimes. The first corresponds to a fast initial relaxation, and the second to the slow barrier hopping. Note that the fast initial relaxation is absent in experiments starting at the cold temperature. $\Delta \mathcal{D}$ is then calculated by fitting the slow regime of the distance curve and interpolating to get the intercept at $t = 0$ (Fig.~\ref{fig:deltaD}a). Thus, $\Delta \mathcal{D}$ is related to $|a_2(\alpha,\,T_\text{initial})|$ by
\begin{align}
	\Delta \mathcal{D} = |a_2(\alpha,\,T_\text{initial})|\, V \,.
\label{Eq:delD}
\end{align}
For the fit based on Eq.~\ref{Eq:a2vsdelD}, each $\mathcal{D}(t)$ decay curve has three parameters, $a_2$, $\lambda_2$, and $V$.  The first, $a_2$, depends on $T_\text{initial}$ and $\alpha$.  Its value differs for each data set.  The other two parameters, $\lambda_2$ and $V$, are common to all the data sets, as they depend only on the properties of the bath.  Thus, the fit is local with respect to  product $|a_2|V$ but global with respect to $\lambda_2$. The fit based on Eq.~\ref{Eq:a2vsdelD} is used to calculate $\Delta \mathcal{D}$ for different $T_\text{initial}$ and $\alpha$ by extrapolating the decay curve back in time to find the intercept at $t = 0$, and subtracting the noise level $\sigma_\mathcal{D}$.

We note that it is also necessary to choose, by hand, the starting point for each decay curve. We verified that the values of the fit parameters are robust to the choice of starting point, typically varying by amounts consistent with the statistical estimates of the parameter error estimates.

After we have extracted  the $\Delta \mathcal{D}$ values from the experiment, we fit to the data a prediction based on Eq.~\ref{Eq:delD}, using $a_2$ coefficients that are numerically calculated from Eq.~\ref{eq:a2}. The remaining fit parameter $V$ agrees with the numerically calculated value based on the eigenfunctions.

For cases where $|a_2(\alpha,T_\text{initial})| \approx 0$, the decay is dominated by $\lambda_3$, and thus, the slow part of the distance curve is absent. In this case, we fit the region of the distance curve after the fast initial kinetic part reaches the noise level.  The result is effectively an upper bound on the size of $a_2$.

\subsection{Calculation of equilibration time}
We define the equilibration time $t_\text{eq}$ to be the time when the distance curve $\mathcal{D}(t)$ reaches the noise level $\sigma_{\mathcal{D}}$.  By equating the terms on the right hand side of Eq.~\ref{Eq:a2vsdelD}, we can determine the time when the slow decay of the distance curve intersects the noise floor, thus reaching equilibrium.  The condition implies
\begin{align}
	\sigma_\mathcal{D} &= |a_2(\alpha,\,T_\text{initial})| \e^{-\lambda_2 t_\text{eq}} V = \Delta \mathcal{D} \e^{-\lambda_2 t_\text{eq}} \,.
\end{align}
Solving for $t_\text{eq}$ gives
\begin{align}
	t_\text{eq} = \frac{1}{\lambda_2} \ln \left[ \frac{\Delta \mathcal{D}}{\sigma_\mathcal{D}}
		\right] \,.
\label{Eq:t_eq}
\end{align}
With $\lambda_2$ and $\Delta \mathcal{D}$ small, normally distributed uncertainties, the variance of $t_\text{eq}$ is
\begin{align}
	\sigma_{t_\text{eq}}^2 \approx 
	\left( \pdv{t_\text{eq}}{\lambda_2} \right)^2 \sigma_{\lambda_2}^2 
		+ \left( \pdv{t_\text{eq}}{\Delta \mathcal{D}} \right)^2 \sigma_{\Delta \mathcal{D}}^2 
		+ 2 \left( \pdv{t_\text{eq}}{\lambda_2} \pdv{f}{\Delta \mathcal{D}}  \right)
			\sigma_{\lambda_2 \Delta \mathcal{D}} \,,
\label{Eq:variance}
\end{align}
where $\sigma_{\lambda_2}$ is the standard deviation of $\lambda_2$, $\sigma_{\Delta \mathcal{D}}$ the standard deviation of $\Delta \mathcal{D}$, and $\sigma_{\lambda_2 \Delta \mathcal{D}}$ the covariance between $\lambda_2$ and $\Delta \mathcal{D}$. Using Eqs.~\ref{Eq:t_eq} and \ref{Eq:variance}, we can write the fractional uncertainty in the equilibration time as
\begin{align}
	\frac{\sigma_{t_\text{eq}}}{t_\text{eq}} 
	= \left[ \left( \frac{\sigma_{\lambda_2}}{\lambda_2}\right)^2
		+ \frac{1}{t_\text{eq}^2\lambda_2^2} \left(\frac{\sigma_{\Delta \mathcal{D}}}
		{\Delta \mathcal{D}}\right)^2-\frac{2}{t_\text{eq}\lambda_2} 
		\left( \frac{\sigma_{\lambda_2 \Delta \mathcal{D}}}{\lambda_2 \Delta \mathcal{D}} \right) 
		\right]^{1/2}.
\label{Eq:SD_t_eq}	
\end{align}
The typical fractional uncertainties in these variables are $(\sigma_{\lambda_2}/\lambda_2) \approx 0.04$, and $(\sigma_{\Delta \mathcal{D}}/\Delta \mathcal{D}) \approx 0.05$, where $\lambda_2 \approx 0.3$ $\text{ms}^{-1}$ and $t_\text{eq}$ varies within the range 1--20 ms.  From Eq.~\ref{Eq:SD_t_eq}, the fractional uncertainty in a typical data for the equilibration time is $(\sigma_{t_\text{eq}}/t_\text{eq}) \approx 0.04$--$0.13$.  Although these are typical numbers, the calculation in Eq.~\ref{Eq:SD_t_eq} is repeated for each data point in Fig.~\ref{eq_time}.  A separate fit performed in each case, with separate fit parameters and parameter uncertainties.  

Finally, because the uncertainty of the noise level $\sigma_\mathcal{D}$ is determined from a long baseline, its fractional value ($\approx 0.006$) is nearly ten times smaller than other fractional uncertainties and does not appreciably alter the uncertainty estimate.  We thus neglect it in our analysis.

\subsection{Equilibration time vs the \texorpdfstring{$a_2$}{a2} coefficient} 
\label{sec: Eq_time_a2}

\begin{figure}[ht!]
\centering
\includegraphics{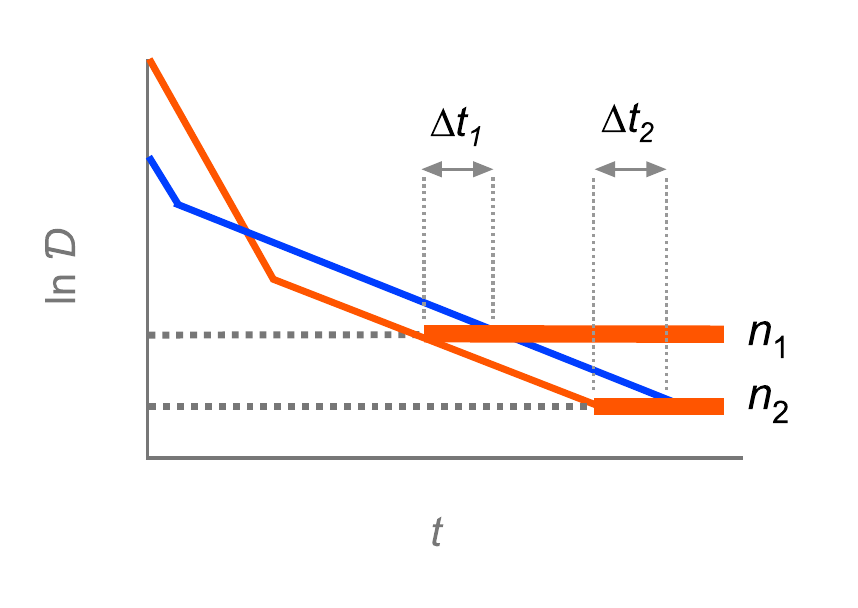}
\caption{\textbf{Different noise levels do not affect the difference in equilibration time.} The hot (red) and warm (blue) systems have the same slope at large times (set by the potential energy).  The signal decreases until it hits one of two different noise levels, $n_1$ or $n_2$ (indicated by thick red lines and horizontal dashes). The difference in the equilibration time is independent of the noise levels:  $\Delta t_1 = \Delta t_2 = t_\text{w}-t_\text{h}$.}
\label{fig:noise}
\end{figure}

To relate the equilibration time to $a_2(\alpha,\,T_\text{initial})$ for the hot and warm cases explicitly, we rewrite Eq.~\ref{Eq:a2vsdelD} at times $t_\text{h}$ and $t_\text{w}$ as
\begin{subequations}
\begin{align}
	\mathcal{D}[p(x,t_\text{w});\pi(x;\,T_\text{b})] 
	&= |a_2(\alpha,\,T_\text{w})| 
		\e^{-\lambda_2 t_\text{w}} V + \sigma_\mathcal{D}, \\
	\mathcal{D}[p(x,t_\text{h});\pi(x;\,T_\text{b})] 
	&=  |a_2(\alpha,\,T_\text{h})| 
		\e^{-\lambda_2 t_\text{h}}V + \sigma_\mathcal{D} \,,
\end{align}
\end{subequations}
where $t_\text{w}$ and $t_\text{h}$ are the equilibration times for the warm and hot systems, respectively.  After both systems have reached equilibrium, the instantaneous value of their  $L_1$ distances from the equilibrium value fluctuate at a typical noise level of $ \sigma_{\mathcal{D}}$ (Eq.~\ref{Eq: L1error}). Equating the two identical average values and simplifying gives 
\begin{align}
	t_\text{w} - t_\text{h} = \frac{1}{\lambda_2} \ln \frac{|a_2(\alpha,\,T_\text{w})|}
		{|a_2(\alpha,\,T_\text{h})|} \,.
\label{time_diff}
\end{align}
Although $t_\text{w}$ and $t_\text{h}$ both increase as the noise level of the distance measure is reduced, their difference is independent of the noise level (Fig.~\ref{fig:noise}). Thus, no matter what the noise level in the estimates of probability densities, we will always reach an unambiguous conclusion concerning the presence of the Mpemba effect.  For $|a_2(\alpha,\,T_\text{w})|> |a_2(\alpha,\,T_\text{h})|$, the warm system lags the hot, and the Mpemba effect is observed. 

\end{document}